\documentclass[a4paper,11pt]{article}
\usepackage[latin1]{inputenc}
\usepackage{indentfirst}
\usepackage{amsfonts}
\usepackage{float}
\usepackage{latexsym,amssymb,amsmath}
\usepackage[dvips]{graphics,graphicx,epsfig,color}
\usepackage[lofdepth,lotdepth]{subfig}
\begin{document}
\thispagestyle{empty}
\title{\bf Dynamical homogenization of a transmission grating}
\author{Armand Wirgin\thanks{LMA, CNRS, UMR 7031, Aix-Marseille Univ, Centrale Marseille, F-13453 Marseille Cedex 13, France, ({\tt wirgin@lma.cnrs-mrs.fr})} }
\date{\today}
\maketitle
\begin{abstract}
  A  periodic assembly of acoustically-rigid blocks (termed 'grating'), situated between two half spaces occupied by fluid-like media, lends itself to a rigorous theoretical analysis  of its  response to an acoustic homogeneous plane wave. This theory  gives rise to two sets of linear equations, the first for the amplitudes of the waves in the space between successive blocks, and the second for the amplitudes of the waves in the two half spaces.  The first set is solved numerically to furnish reference solutions. The second set is submitted to low-frequency approximation procedure whereby the pressure fields  are found to be  those for a  configuration in which the grating becomes a homogeneous layer of the same thickness as the height of the blocks  in the grating. A simple formula is derived for the constitutive properties of this layer in terms of those of the fluid-like medium in between the blocks. The homogeneous layer model scattering amplitude transfer functions  and spectral reflectance, transmittance and absorptance  reproduce  quite well the corresponding rigorous numerical functions of the grating over a non-negligible range of low frequencies.  Due to its simplicity, the homogeneous layer model enables theoretical predictions of  many of the key features of the acoustic response of the grating.
\end{abstract}
Keywords: dynamic response, homogenization, gratings.
\newline
\newline
Abbreviated title: Dynamic homogenization of a grating
\newline
\newline
Corresponding author: Armand Wirgin, \\ e-mail: wirgin@lma.cnrs-mrs.fr
\newpage
\tableofcontents
\newpage
\newpage
\section{Introduction}\label{intro}
The scheme by which  an inhomogeneous (porous, inclusions within a homogeneous host, etc.) medium is reduced (with regard to its response to the solicitation) to a surrogate homogeneous medium is frequently termed  'homogenization'. What is meant by homogenization also includes the manner in which the physical characteristics of the surrogate are related to the structural and physical characteristics of the original, this being  often accomplished by theoretical multiple scattering field averaging techniques \cite{ua49,wt61,sa80,lm06,sp06,tsd06,aa07,cn09} or multiscale techniques (which actually also involve field averaging) \cite{fb05,pa06,mcc17}.  Usually, the latter two techniques cannot be fully-implemented for other than static, or at least low frequency, solicitations \cite{si02,fb05,gpa16}. There exist  alternatives to theoretical  field-averaging and multiscaling, applicable to a range of (usually-low) frequencies,  which can be called:  'computational field averaging' \cite{cm12} and 'computational  parameter retrieval' \cite{wg92,qk11} approaches to dynamic homogenization.

Recent research on metamaterials \cite{la13,lc04,pm04,fb05,fa07,mt10,du12,fp15,rm15,bmm15,bpa15,bua17}  has spurred renewed interest in homogenization techniques \cite{la13,ss02,sv05,fx06,th06,tsd06,aml10,st10,csc13,gr15,ga16,mcc17}, the underlying issue being how to design an inhomogeneous medium so that it responds in a given manner (enhanced absorption \cite{gr15,gpa16,jrg17},  total transmission \cite{ll07}, reduced broadband transmission \cite{lg13,lg16} or various patterns of scattering \cite{wi78, wi88,jc17}) to a wavelike (acoustic \cite{ll07,jrg17}, elastic \cite{pa06}, optical \cite{wi78,wi88}, microwave \cite{km89}, waterwave \cite{ml99})  solicitation, the frequencies of which can exceed the quasi-static regime.  This design problem is, in fact, an inverse problem that is often solved by encasing a specimen of the medium in  a flat layer, treating the latter as if it were homogeneous, and obtaining its constitutive properties (in explicit manner in the NRW technique \cite{nr70,we74,bdp10,ss02,cg04,sv05,go10,la12,la14,wi16b} from the complex amplitudes of the reflected and transmitted waves that constitute the response  of the layer (i.e., the data) to a (normally-incident in the NRW technique) plane body wave. To the question of why encase the specimen within a flat layer  it can be answered that a simple, closed form solution is available for the associated forward problem of the reflected and transmitted-wave response of homogeneous layer to an incident plane body wave (the same being true as regards a stack of horizontal plane-parallel homogeneous layers employed in the geophysical context). Moreover, this solution is valid for all frequencies, layer thicknesses, and even incident angles and polarizations, such as is required in the homogenization problem since one strives to obtain a homogenized medium whose thickness does not depend on the specimen thickness (otherwise, how qualify it as being a medium?), but, on the contrary, he wants to find out how its properties depend on the characteristics of the solicitation. Moreover, the requirement of normal incidence (proper to the NRW technique) can be relaxed provided one accepts to search for the solution of the inverse problem in an optimization, rather than explicit, manner \cite{qk11,wi16b,wi18}.

Herein we shall carry out the homogenization in a more theoretically-oriented manner by means of a three-step procedure: 1) establish the rigorous system of equations for the amplitudes of the reflected and transmitted waves constituting the far-field response of the grating, 2) find a low-frequency approximation of theses amplitudes from the system of equations, 3) show that the so-obtained low-frequency approximate amplitudes correspond exactly to the amplitudes of the reflected and transmitted  waves constituting the response of a homogeneous layer of the same thickness as that of the grating. Moreover, we show, by the same procedure, that the field in the space between blocks of the grating is exactly that of the field in the entire aforementioned homogeneous  layer and we establish the relation of the velocity and mass density of the latter to the velocity and mass density of the filler material as well as to the filling fraction of this material.
\section{Description of the scattering configuration}\label{specif}
The site in which is situated the grating  consists of a lower half space filled with a  homogeneous, isotropic non-lossy  fluid-like medium  wherein the mass density and bulk wave velocity are $\rho^{[0]}$, $c^{[0]}$ (both real) respectively and an upper half space filled with another homogeneous, isotropic fluid-like medium wherein the mass density and bulk wave velocity are $\rho^{[2]}$, $c^{[2]}$ (both real) respectively.

The  grating is composed of a periodic (along $x$, with period $d$) set of identical, perfectly-rigid  rectangular  blocks   whose widths and  heights are $d-w$ and $h$ respectively. The spaces in between successive blocks (width $w$) is filled with a third homogeneous, isotropic generally-lossy fluid-like medium wherein the mass density and bulk wave velocity are $\rho^{[1]}$ (real), $c^{[1]}$ (generally-complex) respectively.

All the geometrical features of the grating and site are assumed to not depend on $y$ and the various interfaces (between the given structure and the two half spaces   are surfaces of constant $z$ and extend indefinitely along $x$ and $y$.

 The solicitation is a homogeneous compressional (i.e., acoustic) plane wave whose  wavevector $\mathbf{k}^{i}$ lies in the $x-z$ (sagittal) plane)), such that $A^{i}(\omega)$ is the spectral amplitude and $\theta^{i}$  the incident angle, with $\omega=2\pi f$ the angular frequency ($f$ the frequency which will be varied).

 The mass densities and bulk wave velocities are assumed to not depend on $f$ (nor, of course on $\theta^{i}$  and $A^{i}$).

\begin{figure}[ht]
\begin{center}
\includegraphics[width=9.0cm] {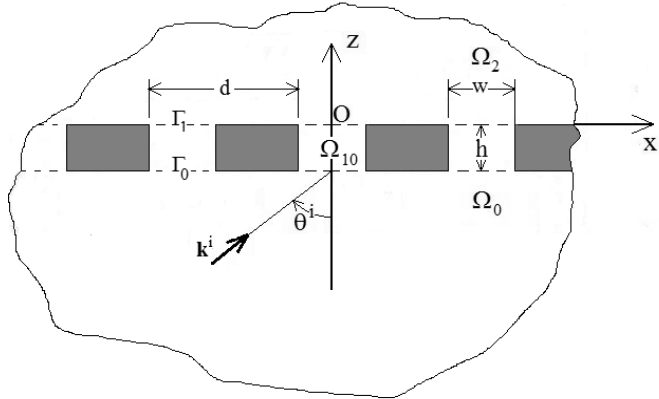}
 \caption{Sagittal plane view of the configuration. $\Omega_{0}$ is the lower half-space domain, $\Omega_{1}=\cup_{n\in\mathbb{Z}}\Omega_{1n}$  the domain occupied by the given periodic structure, $\Omega_{1n}$ the domain of the filler space of width $w$ between successive  blocks,  the blocks being of rectangular cross section  and height $h$, with $d$ the period (along $x$) of the given periodic structure, and $\Omega_{2}$ the  upper half space domain. The  interface (i.e., the line $z=-h_{2}$) between $\Omega_{0}$ and $\Omega_{1}$ is designated by $\Gamma_{0}=\cup_{n\in\mathbb{Z}}\Gamma_{0n}$ and the interface  between  $\Omega_{1}$ and $\Omega_{2}$  by $\Gamma_{1}=\cup_{n\in\mathbb{Z}}\Gamma_{1n}$, in which $\Gamma_{m0}$ is the portion of $\Gamma_{m}$ included between $x=-w/2$ and $x=x/2$.}
  \label{fig1}
  \end{center}
\end{figure}

 The assumed  $y$-independence of the acoustic solicitation, as well as the fact that it was assumed that the blocks of the given structure  do not depend on $y$ entails that all the acoustic field functions do not depend on $y$. Thus, in this 2D   scattering problem, the analysis will take place in the sagittal plane in which the vector joining the origin O to a point whose coordinates are $x,z$ is denoted by $\mathbf{x}$ (see fig. \ref{fig1}.
\clearpage
\newpage
\section{The boundary-value problem}\label{bvp1}
The positive real phase velocities in $\Omega_{l}; l=0,2$ are $c^{[l]}$, and the generally-complex phase velocity in $\Omega_{1n}~n\in \mathbb{Z}$ are $c^{[1]}=c^{'[1]}+ic^{''[1]}$, with $c^{'[1]}>0$ and $c^{''[1]}\le 0$. All $c^{[j]}~l=0,1,2$   are assumed do not depend on $\omega$.

The positive real density in $\Omega_{l}~;~l=0,2$ is $\rho^{[l]}>0$ and the real positive real densities in $\Omega_{1n}~n\in \mathbb{Z}$ are $\rho^{[1]}$.  The three densities are assumed to not depend on $\omega$.

The  wavevector $\mathbf{k}^{i}$  is of the form $\mathbf{k}^{i}=(k_{x}^{i},k_{z}^{i})=(k^{[0]}\sin\theta^{i},k^{[0]}\cos\theta^{i})$ wherein  $\theta^{i}$ is the angle of incidence (see fig. \ref{fig1}), and $k^{[l]}=\omega/c^{[l]}$.

The total pressure wavefield $u(\mathbf{x},\omega)$ in $\Omega_{l}~;~l=0,1,2$ is designated by $u^{[l]}(\mathbf{x},\omega)$.  The incident wavefield is
\begin{equation}\label{1-000}
u^{[0]+}(\mathbf{x},\omega)=u^{i}(\mathbf{x},\omega)=A^{[0]+}(\omega)\exp[i(k_{x}^{i}x+k_{z}^{i}z)]~,
\end{equation}
wherein $A^{[0]+}(\omega)=A^{i}\sigma(\omega)$ and $\sigma(\omega)$ is the spectrum  of the solicitation.

The plane wave nature of the solicitation and the $d$-periodicity of $\Gamma_{0}$ and $\Gamma_{1}$  entails the quasi-periodicity of the field, whose expression is the Floquet condition
\begin{equation}\label{1-005}
u(x+d,z,\omega)=u(x,z,\omega)\exp(ik_{x}^{i}d)~;~\forall\mathbf{x}\in \Omega_{0}+\Omega_{1}+\Omega_{2}~.
\end{equation}
Consequently, as concerns the response in $\Omega_{1}$, it suffices to examine the field in $\Omega_{10}$.

The boundary-value problem in the space-frequency domain translates to the following relations (in which the superscripts $+$ and $-$ refer to the upgoing and downgoing  waves respectively) satisfied by the total displacement field $u^{[l]}(\mathbf{x};\omega)$ in $\Omega_{l}$:
\begin{equation}\label{1-010}
u^{[l]}(\mathbf{x},\omega)=u^{[l]+}(\mathbf{x},\omega)+u^{[l]-}(\mathbf{x},\omega)~;~l=0,1,2~,
\end{equation}
\begin{equation}\label{1-020}
u_{,xx}^{[l]}(\mathbf{x},\omega)+u_{,zz}^{[l]}(\mathbf{x},\omega)+(k^{[l]})^{2}u^{[l]}(\mathbf{x},\omega)=0~;~\mathbf{x}\in \Omega_{l}~;~l=0,1,2~.
\end{equation}
\begin{equation}\label{1-030}
(\rho^{[0]})^{-1}u_{,z}^{[0]}(x,-h,\omega)=0~;~\forall x\in [-d/2,w/2]\cup [w/2,d/2]~,
\end{equation}
\begin{equation}\label{1-033}
(\rho^{[2]})^{-1}u_{,z}^{[2]}(x,0,\omega)=0~;~\forall x\in [-d/2,w/2]\cup [w/2,d/2]~,
\end{equation}
\begin{equation}\label{1-035}
(\rho^{[2]})^{-1}u_{,x}^{[1]}(\pm w/2,z,\omega)=0~;~\forall z\in [0,h_{2}]~,
\end{equation}
\begin{equation}\label{1-040}
u^{[l]}(\mathbf{x},\omega)-u^{[1+1]}(\mathbf{x},\omega)=0~;~\mathbf{x}\in \Gamma_{l}~;~l=0,1~,
\end{equation}
\begin{equation}\label{1-050}
(\rho^{[l]})^{-1}u_{,z}^{[l]}(\mathbf{x},\omega)-(\rho^{[l+1]})^{-1}u_{,z}^{[1+1]}(\mathbf{x},\omega)=0~;~\mathbf{x}\in \Gamma_{l}~;~l=0,1~,
\end{equation}
wherein   $u_{,\zeta}$ ($u_{,\zeta\zeta}$) denotes the first (second) partial derivative of $u$ with respect to $\zeta$. Eq. (\ref{1-020}) is the space-frequency  wave equation for compressional sound waves, (\ref{1-030})-(\ref{1-035}) the rigid-surface  boundary conditions, (\ref{1-040}) the expression of continuity of pressure across the two interfaces $\Gamma_{0}$ and $\Gamma_{0}$  and (\ref{1-050}) the expression of continuity of normal velocity across these same interfaces.

Since $\Omega_{l}~;~l=0,2$ are of half-infinite extent, the field therein must obey the radiation conditions
\begin{equation}\label{1-060}
u^{[l]-}(\mathbf{x},\omega)\sim \text{outgoing waves}~;~\mathbf{x}\rightarrow\infty~;~l=0,2~.
\end{equation}
Various (usually integral equation) rigorous approaches \cite{mi49,km89,ml99} have been employed to solve this boundary value problem. Herein, we outline another rigorous technique, based on the domain decomposition and  separation of variables technique previously developed in \cite{wi78,wi88}.
\subsection{Field representations via separation of variables (DD-SOV)}\label{sov1}
The application of  the domain decomposition-separation of variables (DD-SOV) technique, The Floquet condition, and the radiation conditions gives rise, in the two half-spaces, to the field representations:
\begin{equation}\label{2-010}
u^{[0]}(\mathbf{x},\omega)=\sum_{n\in\mathbb{Z}}\left(A_{0}^{[l]+}(\omega)\exp[i(k_{xn}x+ k_{zn}^{[0]}z)]+
A_{0}^{[l]-}(\omega)\exp[i(k_{xn}x- k_{zn}^{[0]}z)]\right)~,
\end{equation}
\begin{equation}\label{2-011}
u^{[2]}(\mathbf{x},\omega)=\sum_{n\in\mathbb{Z}}A_{n}^{[2]+}(\omega)\exp[i(k_{xn}x+ k_{zn}^{[2]}z)]~,
\end{equation}
wherein:
\begin{equation}\label{2-012}
k_{xn}=k_{x}^{i}+\frac{2n\pi}{d}~,
\end{equation}
\begin{equation}\label{2-020}
k_{zn}^{[l]}=\sqrt{\big(k^{[l]}\big)^{2}-(k_{xn})^{2}}~~;~~\Re k_{zn}^{[l]}\ge 0~~,~~\Im k_{zn}^{[l]}\ge 0~~\omega>0~,
\end{equation}
and, on account of (\ref{1-000}),
\begin{equation}\label{2-030}
A_{n}^{[0]+}(\omega)=A^{[0]+}(\omega)~\delta_{n0}~,
\end{equation}
with $\delta_{n0}$ the Kronecker delta symbol.
In the central block, the DD-SOV technique, together with the rigid body boundary condition (\ref{1-035}), lead to
\begin{equation}\label{2-040}
u^{[1]}(\mathbf{x},\omega)=\sum_{m=0}^{\infty}\left(A_{m}^{[1]+}(\omega))\exp[iK_{zm}^{[1]}z)]+
A_{m}^{[1]-}(\omega))\exp[-iK_{zm}^{[1]}z)]\right)\cos[K_{xm}(x+w/2)]~,
\end{equation}
in which
\begin{equation}\label{2-050}
K_{xm}=\frac{m\pi}{w}~,
\end{equation}
\begin{equation}\label{2-060}
K_{zm}^{[1]}=\sqrt{\big(k^{[1]}\big)^{2}-(K_{xm})^{2}}~~;~~\Re K_{zm}^{[1]}\ge 0~~,~~\Im K_{zm}^{[1]}\ge 0~~\omega>0~.
\end{equation}
%
\subsection{Expressions for each of the four sets of unknowns}
From the remaining boundary and continuity conditions it ensues the four sets of relations:
\begin{equation}\label{4-000}
(\rho^{[0]})^{-1}\int_{-d/2}^{d/2}u_{,z}^{[0]}(x,-h,\omega)\exp(-ik_{xj})\frac{dx}{d}=
(\rho^{[1]})^{-1}\int_{-d/2}^{d/2}u_{,z}^{[1]}(x,-h,\omega)\exp(-ik_{xj})\frac{dx}{d}~;~\forall j\in \mathbb{Z}~,
\end{equation}
\begin{equation}\label{4-005}
(\rho^{[2]})^{-1}\int_{-d/2}^{d/2}u_{,z}^{[2]}(x,0,\omega)\exp(-ik_{xj})\frac{dx}{d}=
(\rho^{[1]})^{-1}\int_{-d/2}^{d/2}u_{,z}^{[1]}(x,0,\omega)\exp(-ik_{xj})\frac{dx}{d}~;~\forall j\in \mathbb{Z}~,
\end{equation}
\begin{multline}\label{4-010}
\int_{-w/2}^{w/2}u^{[0]}(x,-h,\omega)\cos[K_{xl}(x+w/2)]\frac{dx}{w/2}=\\
\int_{-w/2}^{w/2}u^{[1]}(x,-h,\omega)\cos[K_{xl}(x+w/2)]\frac{dx}{w/2}
~;~l=0,1,2,...~,
\end{multline}
\begin{multline}\label{4-015}
\int_{-w/2}^{w/2}u^{[2]}(x,0,\omega)\cos[K_{xl}(x+w/2)]\frac{dx}{w/2}=\\
\int_{-w/2}^{w/2}u^{[1]}(x,0,\omega)\cos[K_{xl}(x+w/2)]\frac{dx}{w/2}
~;~l=0,1,2,...~,
\end{multline}
which should suffice to determine the four sets of unknown coefficients $\{A^{[0]-}\}$, $\{A^{[1]+}\}$, $\{A^{[1]-}\}$, and $\{A^{[2]+}\}$. Employing the DD-SOV field representations in these four relations gives rise to:
\begin{equation}\label{4-020}
A_{j}^{[0]-}=A_{j}^{[0]+}\big(\varepsilon_{j}^{-}\big)^{-2}-
\frac{\rho^{[0]}w}{2\rho^{[1]}d}\frac{\varepsilon_{j}^{-}}{k_{zj}^{[0]}}
\sum_{m=0}^{\infty}\left[A_{m}^{[1]+}e_{m}^{-}-A_{m}^{[1]-}e_{m}^{+}\right]K_{zm}^{[1]}E_{jm}^{-}
~;~\forall j\in\mathbb{Z}~,
\end{equation}
\begin{equation}\label{4-030}
A_{j}^{[2]+}=\frac{\rho^{[2]}w}{2\rho^{[1]}d}\frac{1}{k_{zj}^{[2]}}
\sum_{m=0}^{\infty}\left[A_{m}^{[1]+}-A_{m}^{[1]-}\right]K_{zm}^{[1]}E_{jm}^{-}~;~\forall j\in\mathbb{Z}~,
\end{equation}
\begin{equation}\label{4-040}
A_{l}^{[1]+}e_{l}^{-}+A_{l}^{[1]-}e_{l}^{+}=\frac{\epsilon_{l}}{2}
\sum_{n\in\mathbb{Z}}\left[A_{n}^{[0]+}\varepsilon_{n}^{-}+A_{n}^{[0]-}\varepsilon_{n}^{+}\right]E_{nl}^{+}~;~l=0,1,2,..~,
\end{equation}
\begin{equation}\label{4-050}
A_{l}^{[1]+}+A_{l}^{[1]-}=\frac{\epsilon_{l}}{2}
\sum_{n\in\mathbb{Z}}A_{n}^{[2]+}E_{nl}^{+}~;~l=0,1,2,..~,
\end{equation}
wherein:
\begin{equation}\label{4-060}
e_{m}^{\pm}=\exp(\pm iK_{zm}^{[1]}h)~,~\varepsilon_{n}^{\pm}=\exp(\pm ik_{zn}^{[0]}h)~,~E_{nm}^{\pm}=\int_{-w/2}^{w/2}\exp(\pm ik_{xn}x)\cos[K_{xm}(x+w/2)]\frac{dx}{w/2}
\end{equation}
and $\epsilon_{0}=1$, $\epsilon_{m>0}=2$. More specifically:
\begin{equation}\label{4-070}
E_{nm}^{\pm}=i^{m}\text{sinc}([\pm k_{xn}+K_{xm}]w/2)+i^{-m}\text{sinc}([\pm k_{xn}-K_{xm}]w/2)
\end{equation}
in which $\text{sinc}(\zeta)=\sin(\zeta)/\zeta$ and $\text{sinc}(0)=1$.
\subsection{System of liner equations for the determination of the sets of coefficients $\{A_{m}^{[1]+}\}$ and  $\{A_{m}^{[1]-}\}$}                           %

We can go a step further by inserting two of the expressions (\ref{4-020})-(\ref{4-050}) into the other two and interchange summations to obtain:
\begin{equation}\label{4-080}
\begin{array}{c}
\sum_{m=0}^{\infty}\left(P_{lm}^{11}Q_{m}^{1}+P_{lm}^{12}Q_{m}^{2}\right)=R_{l}^{1}\\\\
\sum_{m=0}^{\infty}\left(P_{lm}^{21}Q_{m}^{1}+P_{lm}^{22}Q_{m}^{2}\right)=R_{l}^{2}
\end{array}
~;~l=0,2,...~,
\end{equation}
in which
\begin{equation}\label{4-090}
P_{lm}^{11}=\delta_{lm}+\frac{\epsilon_{l}w\rho^{[0]}}{4d\rho^{[1]}}e_{l}^{+}e_{m}^{-}K_{zm}^{[1]}S^{[0]}_{lm}~,~
P_{lm}^{12}=\left(e_{l}^{+}\right)^{2}\delta_{lm}-\frac{\epsilon_{l}w\rho^{[0]}}{4d\rho^{[1]}}e_{l}^{+}e_{m}^{+}K_{zm}^{[1]}S^{[0]}_{lm}
~,~R_{m}^{1}=A^{[0]+}\epsilon_{l}e_{l}^{+}\varepsilon_{0}^{-}E_{0l}^{+}
\end{equation}
\begin{equation}\label{4-100}
P_{lm}^{21}=\delta_{lm}-\frac{\epsilon_{l}w\rho^{[2]}}{4d\rho^{[1]}}e_{l}^{+}e_{m}^{-}K_{zm}^{[1]}S^{[2]}_{lm}~,~
P_{lm}^{22}=\delta_{lm}+\frac{\epsilon_{l}w\rho^{[2]}}{4d\rho^{[1]}}K_{zm}^{[1]}S^{[0]}_{lm}
~,~R_{m}^{2}=0~,~S_{lm}^{[j]}=\sum_{n\in\mathbb{Z}}\frac{E_{nl}^{+}E_{nm}^{-}}{k_{zn}^{[j]}}~,
\end{equation}
and $Q_{m}^{1}=A_{m}^{[1]+}$, $Q_{m}^{2}=A_{m}^{[1]-}$.
\subsection{Numerical issues concerning the resolution of the system of equations for $\{A_{m}^{[1]\pm}\}$}
We strive to obtain numerically the sets $\{A_{m}^{[1]\pm}\}$ from the linear system of equations (\ref{4-080}). Once these sets are found, they are introduced into (\ref{4-020})-(\ref{4-030}) to obtain the sets $\{A_{n}^{[0]-}\}$ and $\{A_{n}^{[2]+}\}$.

Concerning the resolution of the infinite system of linear equations (\ref{4-080}), the procedure is basically to replace it by the finite system of linear equations
\begin{equation}\label{5-010}
\begin{array}{c}
\sum_{m=0}^{M}\left(P_{lm}^{11(N)}Q_{m}^{1}+P_{lm}^{12(N)}Q_{m}^{2}\right)=R_{l}^{1}\\\\
\sum_{m=0}^{M}\left(P_{lm}^{21(N)}Q_{m}^{1}+P_{lm}^{22(N)}Q_{m}^{2}\right)=R_{l}^{2}
\end{array}
~;~l=0,2,...,M~,
\end{equation}
in which $P_{lm}^{jk(N)}$ signifies that the series in $S_{lm}^{[k]}$ therein is limited to the terms $n=0,\pm 1,...,\pm N$, $N$ having been chosen to be sufficiently large to obtain numerical convergence of these series, and to increase $M$ so as to generate the sequence of numerical solutions $\{Q_{0}^{j(0)}\}$, $\{Q_{0}^{j(1)},Q_{1}^{j(1)}\}$,....until the values of the first few members of these sets stabilize and the remaining members become very small. This is usually obtained for  values of $M$, that are all the smaller  the lower is the  frequency.

When all the coefficients (we mean those whose values depart significantly from zero) are found, they enable the computation of the far-field and near-field acoustic responses  that are of interest herein. The so-obtained numerical solutions for these  coefficients and fields, can, for all practical purposes,  be considered to be 'exact' since they compare very well with their finite element or integral equation counterparts in  \cite{mi49,km89, ml99}.
\subsection{System of liner equations for the determination of the sets of coefficients $\{A_{n}^{[0]-}\}$ and  $\{A_{n}^{[2]+}\}$}                           %
We can proceed differently by inserting (\ref{4-040})-(\ref{4-050}) into  (\ref{4-020})-(\ref{4-030})  and interchanging summations to obtain:
\begin{equation}\label{4-200}
\begin{array}{c}
\sum_{n\in\mathbb{Z}}\left(X_{jn}^{11}Y_{n}^{1}+X_{jn}^{12}Y_{n}^{2}\right)=Z^{1}_{j}\\\\
\sum_{n\in\mathbb{Z}}\left(X_{jn}^{21}Y_{n}^{1}+X_{jn}^{22}Y_{n}^{2}\right)=Z^{2}_{j}
\end{array}
~;~j\in\mathbb{Z}~,
\end{equation}
in which
\begin{equation}\label{4-210}
X_{jn}^{11}=\delta_{jn}-\frac{\rho^{[0]}w}{\rho^{[1]}4id}\frac{\varepsilon_j^{-}}{k_{zj}^{[0]}}\varepsilon_{n}^{+}\Sigma^{2}_{jn}~,~
X_{jn}^{12}=\frac{\rho^{[0]}w}{\rho^{[1]}4id}\frac{\varepsilon_{j}^{-}}{k_{zj}^{[0]}}\Sigma^{1}_{jn}~,~
Z_{j}^{1}=A_{j}^{[0]+}\left(\varepsilon_{j}^{-}\right)^{2}+\frac{\rho^{[0]}w}{\rho^{[1]}4id}\frac{\varepsilon_{j}^{-}}{k_{zj}^{[0]}}
\sum_{\nu\in\mathbb{Z}}A_{\nu}^{[0]+}\varepsilon_{\nu}^{-}\Sigma_{j\nu}^{2}
\end{equation}
\begin{equation}\label{4-220}
X_{jn}^{21}=\frac{\rho^{[2]}w}{\rho^{[1]}4id}\frac{1}{k_{zj}^{[2]}}\varepsilon_{n}^{+}\Sigma^{1}_{jn}~,~
X_{jn}^{22}=\delta_{lm}-\frac{\rho^{[2]}w}{\rho^{[1]}4id}\frac{1}{k_{zj}^{[2]}}\Sigma^{2}_{jn}
~,~Z_{j}^{2}=-\frac{\rho^{[2]}w}{\rho^{[1]}4id}\frac{1}{k_{zj}^{[2]}}
\sum_{\nu\in\mathbb{Z}}A_{\nu}^{[0]+}\varepsilon_{\nu}^{-}\Sigma_{j\nu}^{1}~,
\end{equation}
\begin{equation}\label{4-230}
\Sigma_{jn}^{1}=\sum_{m=0}^{\infty}\epsilon_{m}\frac{K_{zm}^{[1]}}{S_{m}}E_{jm}^{-}E_{nm}^{+}~,~
\Sigma_{jn}^{2}=\sum_{m=0}^{\infty}\epsilon_{m}\frac{K_{zm}^{[1]}C_{m}}{S_{m}}E_{jm}^{-}E_{nm}^{+}~,
\end{equation}
and
\begin{equation}\label{4-240}
Y_{n}^{1}=A_{n}^{[0]-},~~Y_{n}^{2}=A_{m}^{[2]+},~~C_{m}=\cos(K_{zm}^{[1]}h),~~S_{m}=\sin(K_{zm}^{[1]}h)~.
\end{equation}
%
\section{Approximate solution for the acoustic response of the transmission grating}
We now discuss an iterative approach for the resolution of the second system of equations for  $\{A_{n}^{[0]-}\}$ and  $\{A_{n}^{[2]+}\}$. We shall limit ourselves to the first-order iterate.
\subsection{The iterative scheme}
 We can write (\ref{4-200}) as
\begin{equation}\label{4-250}
\begin{array}{c}
X_{jj}^{11}Y_{j}^{1}+X_{jj}^{12}Y_{j}^{2}=Z^{1}_{j}-\sum_{n=-\infty,\ne j}^{\infty}\left(X_{jn}^{11}Y_{n}^{1}+X_{jn}^{12}Y_{n}^{2}\right)\\\\
X_{jj}^{21}Y_{j}^{1}+X_{jj}^{22}Y_{j}^{2}=Z^{2}_{j}-\sum_{n=-\infty,\ne j}^{\infty}\left(X_{jn}^{21}Y_{n}^{1}+X_{jn}^{22}Y_{n}^{2}\right)
\end{array}
~;~j\in\mathbb{Z}~.
\end{equation}
 It ensues formally:
\begin{equation}\label{4-260}
Y_{j}^{1}=\frac{W_{j}^{1}X_{jj}^{22}-W_{j}^{2}X_{jj}^{12}}{X_{jj}^{11}X_{jj}^{22}-X_{j}^{21}X_{jj}^{12}}~,~
Y_{j}^{2}=\frac{W_{j}^{2}X_{jj}^{11}-W_{j}^{1}X_{jj}^{21}}{X_{jj}^{11}X_{jj}^{22}-X_{j}^{21}X_{jj}^{12}}
~;~j\in\mathbb{Z}~,
\end{equation}
wherein
\begin{equation}\label{4-270}
W_{j}^{1}=Z^{1}_{j}-\sum_{n=-\infty,\ne j}^{\infty}\left(X_{jn}^{11}Y_{n}^{1}+X_{jn}^{12}Y_{n}^{2}\right)~,~\\
W_{j}^{2}=Z^{2}_{j}-\sum_{n=-\infty,\ne j}^{\infty}\left(X_{jn}^{21}Y_{n}^{1}+X_{jn}^{22}Y_{n}^{2}\right)~,~
~.
\end{equation}
The iterative scheme is then;
\begin{equation}\label{4-280}
Y_{j}^{1(k)}=\frac{W_{j}^{1(k-1)}X_{jj}^{22}-W_{j}^{2(k-1)}X_{jj}^{12}}{D_{j}}~,~
Y_{j}^{2(k)}=\frac{W_{j}^{2(k-1)}X_{jj}^{11}-W_{j}^{1(k-1)}X_{jj}^{21}}{D_{j}}
~;~j\in\mathbb{Z}~,~k=1,2,...~,
\end{equation}
wherein
\begin{equation}\label{4-285}
D_{j}=X_{jj}^{11}X_{jj}^{22}-X_{j}^{21}X_{jj}^{12}~.
\end{equation}
and
\begin{multline}\label{4-290}
W_{j}^{1(k-1)}=Z^{1}_{j}-\sum_{n=-\infty,\ne j}^{\infty}\left(X_{jn}^{11}Y_{n}^{1(k-1)}+X_{jn}^{12}Y_{n}^{2(k-1)}\right)~,~\\
W_{j}^{2(k-1)}=Z^{2}_{j}-\sum_{n=-\infty,\ne j}^{\infty}\left(X_{jn}^{21}Y_{n}^{1(k-1)}+X_{jn}^{22}Y_{n}^{2(k-1)}\right)~,~
~.
\end{multline}
In these expressions, we have not yet addressed the question of zeroth-order  iterates $Y_{j}^{1(0)}$, $Y_{j}^{2(0)}$.
\subsection{The low-order iterates in a low-frequency, large filling factor context}
The preceding formulae show that we must initiate the iteration procedure with a priori assumptions concerning $Y_{j}^{1(0)}$, $Y_{j}^{2(0)}$. We place ourselves in   situations characterized by
\begin{equation}\label{6-100}
k^{[0]}<|k_{xj}|~;~\forall|j|>0~,
\end{equation}
\begin{equation}\label{6-101}
k^{[2]}<|k_{xj}|~;~\forall|j|>0~,
\end{equation}
which specify that only the $j=0$  waves in the bottom and top media are homogeneous. In addition, we assume that under these conditions, all the $j\ne 0$ (inhomogeneous) waves have vanishing amplitudes, which translates to
\begin{equation}\label{66-110}
Y_{j\ne 0}^{1(0)}=Y_{j\ne 0}^{2(0)}=0~.
\end{equation}
We also shall assume
\begin{equation}\label{6-090}
k^{[0]}w<<\pi~\Rightarrow~|k_{x}^{i}|w<<\pi~,
\end{equation}
which together  with (\ref{6-100}),  (\ref{6-101}) and (\ref{66-110}) corresponds to an essentially low-frequency context.

We make one further assumption
\begin{equation}\label{6-103}
w/d\approx 1~,
\end{equation}
which corresponds to a large filling fraction (i.e., narrow block) situation.

Let us now see what the consequences of these assumptions are. We had
\begin{multline}\label{6-097}
E_{0m}^{\pm}=
i^{m}\{\text{sinc}[(\pm k_{x0}+K_{xm})w/2]+(-1)^{m}\text{sinc}[(\pm k_{x0}-K_{xm})w/2]\}=\\
i^{m}\{\text{sinc}[(\pm k_{x}^{i}w/2+m\pi/2]+(-1)^{m}\text{sinc}[(\pm k_{x}^{i}w/2-m\pi/2]\}
~,
\end{multline}
or, on account of (\ref{6-090}),
\begin{equation}\label{6-099}
E_{0m}^{\pm}\approx i^{m}[1+(-1)^{m}]\text{sinc}(m\pi/2)=2\delta_{m0}~.
\end{equation}
Furthermore, (\ref{6-090}) and  (\ref{6-103}) tell us that
\begin{equation}\label{6-104}
E_{n0}^{\pm}=i^{m}\{1+(-1)^{m}\}\text{sinc}(\pm n\pi w/d)\approx 2\text{sinc}(n\pi)=2\delta_{n0}
~,
\end{equation}
the consequences of which are (via (\ref{4-230})):
\begin{equation}\label{66-120}
\Sigma_{j0}^{1}=4\frac{K_{z0}^{[1]}}{S_{0}}\delta_{j0}=\Sigma_{00}^{1}\delta_{j0}~~,
~~\Sigma_{j0}^{2}=4\frac{K_{z0}^{[1]}C_{0}}{S_{0}}\delta_{j0}=\Sigma_{00}^{2}\delta_{j0}
~,
\end{equation}
It then follows from (\ref{66-110}) ad  (\ref{66-120}) that:
\begin{equation}\label{66-130}
W_{j}^{1(0)}=Z_{0}^{1}\delta_{j0}~~,~~W_{j}^{2(0)}=Z_{0}^{2}\delta_{j0}
~,
\end{equation}
whence
\begin{multline}\label{4-300}
Y_{j}^{1(1)}=\frac{W_{j}^{1(0)}X_{jj}^{22}-W_{j}^{2(0)}X_{jj}^{12}}{D_{j}}=
\left(\frac{Z_{0}^{1}X_{00}^{22}-Z_{0}^{2}X_{00}^{12}}{D_{0}}\right)\delta_{j0}=
Y_{0}^{1(1)}\delta_{j0}=\frac{N_{0}^{1(1)}}{D_{0}}\delta_{j0}\\
Y_{j}^{2(1)}=\frac{W_{j}^{2(0)}X_{jj}^{11}-W_{j}^{1(0)}X_{jj}^{21}}{D_{j}}=
\left(\frac{Z_{0}^{2}X_{00}^{11}-Z_{0}^{1}X_{00}^{21}}{D_{0}}\right)\delta_{j0}=
Y_{0}^{2(1)}\delta_{j0}=\frac{N_{0}^{2(1)}}{D_{0}}\delta_{j0}~.
\end{multline}
Eq. (\ref{66-120}) entails

\begin{equation}\label{4-320}
X_{00}^{11}=1-\frac{\rho^{[0]}k_{z0}^{[1]}w}{\rho^{[1]}k_{z0}^{[0]}d}\frac{C_{0}}{iS_{0}}~,~
X_{00}^{12}=\frac{\rho^{[0]}k_{z0}^{[1]}w}{\rho^{[1]}k_{z0}^{[0]}d}\frac{\varepsilon_{0}^{-}}{iS_{0}}~,~
X_{00}^{21}=\frac{\rho^{[2]}k_{z0}^{[1]}w}{\rho^{[1]}k_{z0}^{[2]}d}\frac{\varepsilon_{0}^{+}}{iS_{0}}~,~
X_{00}^{22}=1-\frac{\rho^{[2]}k_{z0}^{[1]}w}{\rho^{[1]}k_{z0}^{[2]}d}\frac{C_{0}}{iS_{0}}~.
\end{equation}
Consequently:
\begin{multline}\label{4-330}
D_{0}=X_{00}^{11}X_{00}^{22}-X_{00}^{21}X_{00}^{12}=\\
\left(1-\frac{\rho^{[0]}k_{z0}^{[1]}w}{\rho^{[1]}k_{z0}^{[0]}d}\frac{C_{0}}{iS_{0}}\right)
\left(1-\frac{\rho^{[2]}k_{z0}^{[1]}w}{\rho^{[1]}k_{z0}^{[2]}d}\frac{C_{0}}{iS_{0}}\right)-
\left(\frac{\rho^{[2]}k_{z0}^{[1]}w}{\rho^{[1]}k_{z0}^{[2]}d}\frac{\varepsilon_{0}^{+}}{iS_{0}}\right)
\left(\frac{\rho^{[0]}k_{z0}^{[1]}w}{\rho^{[1]}k_{z0}^{[0]}d}\frac{\varepsilon_{0}^{-}}{iS_{0}}\right)
~.
\end{multline}
This expression is easily shown to reduce to:
\begin{equation}\label{4-340}
D_{0}=\left(\frac{i}{g_{0}g_{2}S_{0}}\right)\left[g_{1}\big(g_{2}+g_{0}\big)C_{0}-\big(g_{2}g_{0}+g_{1}g_{1}\big)iS_{0}\right]~,
\end{equation}
in which
\begin{equation}\label{4-350}
g_{0}=\frac{k_{z0}^{[0]}}{\rho^{[0]}}~,~g_{1}=\frac{k_{z0}^{[1]}w}{\rho^{[1]}d}~,~g_{2}=\frac{k_{z0}^{[2]}}{\rho^{[2]}}~.
\end{equation}
By the same token,
\begin{equation}\label{4-370}
Z_{0}^{1}=A^{[0]+}\left(\varepsilon_{0}^{-}\right)^{2}\left[1+\frac{g_{1}C_{0}}{g_{0}iS_{0}}\right]~~,~~
Z_{0}^{2}=-A^{[0]+}\varepsilon_{0}^{-}\left[1+\frac{g_{1}}{g_{2}iS_{0}}\right]
~,
\end{equation}
so   that:
\begin{multline}\label{4-380}
N_{0}^{1(1)}=Z_{0}^{1}X_{00}^{22}-Z_{0}^{2}X_{00}^{12}=
A^{[0]+}\left(\varepsilon_{0}^{-}\right)^{2}\left[1+\frac{g_{1}C_{0}}{g_{0}iS_{0}}\right]\left[1-\frac{g_{1}C_{0}}{g_{2}iS_{0}}\right]+
A^{[0]+}\varepsilon_{0}^{-}\left[\frac{g_{1}}{g_{2}iS_{0}}\right]\left[\frac{g_{1}\varepsilon_{0}^{-}}{g_{0}iS_{0}}\right]~~,\\
N_{0}^{2(1)}=Z_{0}^{2}X_{00}^{11}-Z_{0}^{1}X_{00}^{21}=
-A^{[0]+}\varepsilon_{0}^{-}\left[\frac{g_{1}}{g_{2}iS_{0}}\right]\left[1-\frac{g_{1}C_{0}}{g_{0}iS_{0}}\right]-
A^{[0]+}\left(\varepsilon_{0}^{-}\right)^{2}\left[1+\frac{g_{1}C_{0}}{g_{0}iS_{0}}\right]\left[\frac{g_{1}\varepsilon_{0}^{+}}{g_{2}iS_{0}}\right]~,
\end{multline}
which reduce to:
\begin{equation}\label{4-390}
N_{0}^{1(1)}=-A^{[0]+}\left(\varepsilon_{0}^{-}\right)^{2}\frac{i}{g_{0}g_{2}}
\left[-\big(g_{1}g_{1}-g_{2}g_{0}\big)iS_{0}+g_{1}\big(g_{2}-g_{0}\big)C_{0}\right]~,~
N_{0}^{2(1)}=-A^{[0]+}\varepsilon_{0}^{-}\left[\frac{2g_{1}}{g_{2}iS_{0}}\right]~,
\end{equation}
whence:
\begin{multline}\label{4-400}
Y_{0}^{1(1)}=A_{0}^{[0]-(1)}=-A^{[0]+}\left(\varepsilon_{0}^{-}\right)^{2}\left[\frac{g_{1}(g_{2}-g_{0})C_{0}+(g_{2}g_{0}-g_{1}g_{1})iS_{0}}
{g_{1}(g_{2}+g_{0})C_{0}-(g_{2}g_{0}+g_{1}g_{1})iS_{0}}\right]~,\\
Y_{0}^{2(1)}=A_{0}^{[2]+}=A^{[0]+}\varepsilon_{0}^{-}\left[\frac{2g_{1}g_{0}}
{g_{1}(g_{2}+g_{0})C_{0}-(g_{2}g_{0}+g_{1}g_{1})iS_{0}}\right]~,
\end{multline}
and we recall (\ref{4-300})
\begin{equation}\label{4-410}
Y_{j\ne 0}^{1(1)}=A_{j\ne 0}^{[0]-(1)}=0~~,~~Y_{j\ne 0}^{2(1)}=A_{j\ne 0}^{[2]+(1)}=0 ~.
\end{equation}
We now turn to the first-order iterate of the coefficients of the field in the region between successive blocks. From (\ref{4-040})-(\ref{4-050}) we obtain
\begin{equation}\label{4-440}
A_{l}^{[1]+(1)}e_{l}^{-}+A_{l}^{[1]-(1)}e_{l}^{+}=\frac{\epsilon_{l}}{2}
\sum_{n\in\mathbb{Z}}\left[A_{n}^{[0]+}\varepsilon_{n}^{-}+A_{n}^{[0]-(1)}\varepsilon_{n}^{+}\right]E_{nl}^{+}~;~l=0,1,2,..~,
\end{equation}
\begin{equation}\label{4-450}
A_{l}^{[1]+(1)}+A_{l}^{[1]-(1)}=\frac{\epsilon_{l}}{2}
\sum_{n\in\mathbb{Z}}A_{n}^{[2]+(1)}E_{nl}^{+}~;~l=0,1,2,..~,
\end{equation}
Using the (\ref{4-400})-(\ref{4-410}) gives rise (with the help of (\ref{6-099})) to
\begin{equation}\label{4-470}
A_{l}^{[1]+(1)}e_{l}^{-}+A_{l}^{[1]-(1)}e_{l}^{+}=\frac{\epsilon_{l}}{2}
\left[A_{0}^{[0]+}\varepsilon_{0}^{-}+A_{0}^{[0]-(1)}\varepsilon_{0}^{+}\right]E_{0l}^{+}=
\left[A_{0}^{[0]+}\varepsilon_{0}^{-}+A_{0}^{[0]-(1)}\varepsilon_{0}^{+}\right]\delta_{l0}~;~l=0,1,2,..~,
\end{equation}
\begin{equation}\label{4-480}
A_{l}^{[1]+(1)}+A_{l}^{[1]-(1)}=\frac{\epsilon_{l}}{2}
A_{0}^{[2]+(1)}E_{0l}^{+}~;~l=A_{0}^{[2]+(1)}\delta_{l0}~;~l=0,1,2,..~,
\end{equation}
The introduction of (\ref{4-400})-(\ref{4-410}) into (\ref{4-470})-(\ref{4-480}) finally yields
\begin{multline}\label{4-480}
A_{l}^{[1]-(1)}=A_{0}^{+}\varepsilon_{0}^{-}\left[\frac{g_{0}(g_{1}-g_{2})}
{g_{1}(g_{2}+g_{0})C_{0}-(g_{2}g_{0}+g_{1}g_{1})i S_{0}}\right]\delta_{l0}~,~\\
A_{l}^{[1]+(1)}=A_{0}^{+}\varepsilon_{0}^{-}\left[\frac{g_{0}(g_{1}+g_{2})}
{g_{1}(g_{2}+g_{0})C_{0}-(g_{2}g_{0}+g_{1}g_{1})i S_{0}}\right]\delta_{l0}~;~l=0,1,2,..~.
\end{multline}
The higher-order iterates of the various field coefficients can be obtained in similar manner, but we shall content ourselves with the first order results. However, the latter will be compared to the solutions of a homogeneous layer scattering problem to see what are the connections, if any, between the two.

A last word is here in order  about the fields associated with the first-order approximation of the coefficients. These fields are:
\begin{multline}\label{4-480}
u^{[0](1)}=u^{[0]+}+
\sum_{n\in\mathbb{Z}}A_{n}^{[0]-(1)}\exp[i(k_{xn}x-k_{zn}^{[0]}z)]=\\
A_{0}^{+}\exp[i(k_{x0}x+k_{z0}^{[0]}z)]+A_{0}^{[0]-(1)}\exp[i(k_{x0}x-k_{z0}^{[0]}z)]~,
\end{multline}
\begin{multline}\label{4-490}
u^{[1](1)}=
\sum_{m=0}^{\infty}\left[
A_{m}^{[1]+(1)}\exp(i K_{zm}^{[1]}z)+
A_{m}^{[1]-(1)}\exp(-iK_{zm}^{[1]}z)
\right]
\cos[K_{xm}(x+w/2)]=\\
A_{0}^{[1]+(1)}\exp(i K_{z0}^{[1]}z)+A_{0}^{[1]-(1)}\exp(-iK_{z0}^{[1]}z)]~,
\end{multline}
\begin{equation}\label{4-500}
u^{[2](1)}=\sum_{n\in\mathbb{Z}}A_{n}^{[2]+(1)}\exp[i(k_{xn}x+k_{zn}^{[2]}z)]=
A_{0}^{[2]+(1)}\exp[i(k_{x0}x+k_{z0}^{[2]}z)]~,
\end{equation}
%
\subsection{A conservation principle for the grating}\label{consv}
Using Green's second identity, it is rather straightforward to demonstrate the following conservation principle:
\begin{equation}\label{5-020}
\rho+\alpha+\tau=1~,
\end{equation}
wherein  $\rho$ an and $\tau$ are what can be termed hemispherical reflected and transmitted fluxes respectively,  given by:
\begin{equation}\label{5-030}
\rho=\Re\sum_{n\in\mathbb{Z}}\frac{\|A_{n}^{[0]-}\|^{2}k_{zn}^{[0]}}{\|A^{[0]+}\|^{2}k_{z0}^{[0]}}~~,~~
\tau=\Re\sum_{n\in\mathbb{Z}}\frac{\|A_{n}^{[2]-}\|^{2}k_{zn}^{[2]}\rho^{[0]}}{\|A^{[0]+}\|^{2}k_{z0}^{[0]}\rho^{[2]}}~
\end{equation}
and $\alpha$ is what can be called the absorbed flux given by
\begin{equation}\label{5-035}
\alpha=\frac{1}{k_{z0}^{[0]}d}\frac{\rho^{[0]}}{\rho^{[1]}}
\Im\left[\left(k^{[1]}\right)^{2}\right]\int_{\Omega_{10}}\left\|\frac{u^{[1]}(\mathbf{x},\omega)}{A^{[0]+}(\omega)}\right\|^{2}d\varpi~,
\end{equation}
wherein $d\varpi$ is the infinitesimal element of area in the central interstitial domain $\Omega_{10}$ of the grating, it having been assumed, as previously, that the media filling both the lower and upper half-spaces are non-lossy. In addition,  (\ref{5-035}) tells us that the absorbed flux vanishes when the medium filling the interstitial spaces is non-lossy (i.e. when $c^{[1]}$, and therefore $k^{[1]}$, are real.

The expression (\ref{5-030}) shows (since $k_{z0}^{[0]}=k_{z}^{i}$ is real) that the only contributions to $\rho$ stem from diffracted-reflected waves for which $k_{zn}^{[0]}$ is real, and the only contributions to $\tau$ stem from diffracted-transmitted waves for which $k_{zn}^{[2]}$ is real. Real $k_{zn}^{[0]}$ corresponds to homogeneous  plane waves in the lower half space and real $k_{zn}^{[2]}$ to homogeneous  plane waves in the upper half space. The angles of emergence of these observable homogeneous waves are $\theta_{n}^{-}=\arcsin(k_{xn}/k^{[0]})$ and $\theta_{n}^{+}=\arcsin(k_{xn}/k^{[2]})$ (note that $\theta_{0}^{-}=\theta^{i}$ is the angle of specular reflection in the sense of Snell, and $\theta_{0}^{+}=\arcsin(k^{[0]}\sin\theta^{i}/k^{[2]})$ the angle of refraction in the sense of Fresnel). Thus, the flux in each half space is composed of a denumerable, finite, set of  subfluxes, which fact is expressed by:
\begin{equation}\label{5-040}
\rho=\sum_{n\in\mathbb{H}^{-}}\rho_{n}~~,~~
\tau=\sum_{n\in\mathbb{H}^{+}}\tau_{n}~.
\end{equation}
wherein:
\begin{equation}\label{5-050}
 \rho_{n}^{-}=\frac{\|A_{n}^{[0]-}\|^{2}k_{zn}^{[0]}}{\|A^{[0]+}\|^{2}k_{z0}^{[0]}}~~,~~
 \tau_{n}^{+}=\frac{\|A_{n}^{[2]-}\|^{2}k_{zn}^{[2]}\rho^{[0]}}{\|A^{[0]+}\|^{2}k_{z0}^{[0]}\rho^{[2]}}~.
\end{equation}
and $\mathbb{H}^{-}$  the set of $n$ for which $k_{zn}^{[0]}$ is real, whereas  $\mathbb{H}^{+}$ is the set of $n$ for which $k_{zn}^{[2]}$ is real.

Actually, (\ref{5-020}) can be resumed by the following conservation relation
\begin{equation}\label{5-055}
f_{out}=f_{in}~,
\end{equation}
wherein $f_{out}=\rho+\alpha+\tau$ is the normalized output flux and $f_{in}=1$ the normalized input flux.

It is important to underline the fact that the conservation principle is a rigorous consequence of the grating boundary-value problem and therefore does not depend on the low-frequency, large filling fraction approximations made previously. However, this principle can (and will) be employed to test the consistency of both the rigorous and first-iterate solutions of the grating response.
\section{Acoustic response of  a homogeneous layer situated between two fluid-like half spaces}
\begin{figure}[ptb]
\begin{center}
\includegraphics[width=10cm] {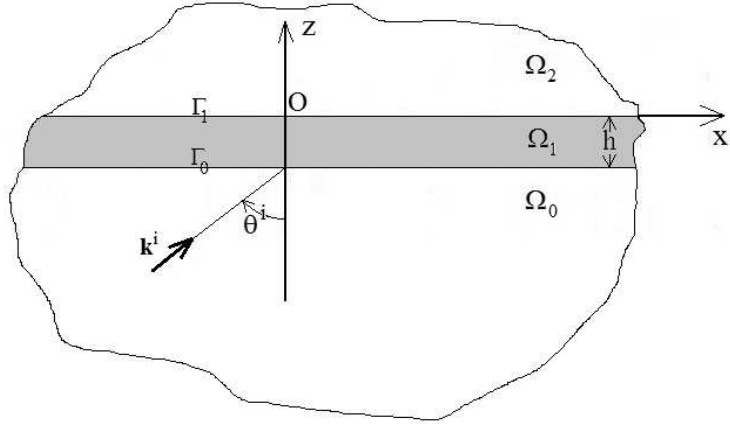}
 \caption{Sagittal plane view of the homogeneous layer configuration. $\Omega_{0}$ is the lower  half-space domain,  $\Omega_{1}$  the  surrogate  domain in the form of a layer  of thickness $h$. $\Omega_{2}$ is the half-space above the superstructure. The  interface (i.e., the line $z=-h$) between $\Omega_{0}$ and $\Omega_{1}$ is designated by $\Gamma_{0}$  and the interface  between  $\Omega_{1}$ and $\Omega_{2}$  by $\Gamma_{1}$.}
  \label{fig2}
  \end{center}
\end{figure}

This configuration, depicted in fig. \ref{fig2}, is similar to the previous one, except that the grating is replaced by a homogeneous layer situated between the same two half spaces. Its acoustic  response can be treated in the classical manner \cite{ejp57} outlined hereafter.

The solicitation, bottom and top half spaces are as previously (i.e., in the problem corresponding to the grating). The surrogate occupies the layer-like domain $\Omega_{1}$ (see fig. \ref{fig2}) in which the properties are designated by the superscript 1. Thus, the boundary-value problem is expressed by (\ref{1-000}), (\ref{1-010}), (\ref{1-020}), (\ref{1-040}), (\ref{1-050}), (\ref{1-060}) (in which $\rho^{[l]}$ is replaced by $R^{[l]}$),  $c^{[l]}$  by $C^{[l]}$),  and $u^{[l]}$  by $U^{[l]}$), with the understanding that:
\begin{equation}\label{7-015}
C^{[l]}=c^{[l]}~~,~~R^{[l]}=\rho^{[l]}~;~l=0,2~.
\end{equation}
\begin{equation}\label{7-015}
U^{[0]+}(\mathbf{x},\omega)=u^{[0]+}(\mathbf{x},\omega)=u^{i}(\mathbf{x},\omega)=A^{[0]+}\exp[i(k_{x}^{i}x+k_{z}^{i}z]~,
\end{equation}
with $k_{x}^{i}=K^{[0]}\sin\theta^{i}$, $k_{z}^{i}=K^{[0]}\cos\theta^{i}$ and $K^{[l]}=\omega/C^{[l]}~;~l=0,1,2$.
\subsection{DD-SOV Field representations}\label{sov2}
Separation of variables and the radiation condition lead to the field representations:
\begin{equation}\label{7-017}
U^{[l]}(\mathbf{x},\omega)=U^{[l]+}(\mathbf{x},\omega)+U^{[l]-}(\mathbf{x},\omega)~;~\forall x\in \Omega_{l}~,~l=0,1,2~.
\end{equation}
with
\begin{equation}\label{7-020}
U^{[l]\pm}(\mathbf{x},\omega)=A^{[l]\pm}\exp\left[i(k_{x}x\pm k_{z}^{[l]}z)\right]~;~\forall x\in \Omega_{l}~,~l=0,1,2~.
\end{equation}
  in which $A^{[0]+}$ is as previously, and the relation to previous wavenumbers is as follows:  $k_{z}^{[l]}=k_{z0}^{[l]}=\sqrt{\big(K^{[l]}\big)^2-\big(k_{x}\big)^2}$.
\subsection{Solutions for the plane-wave coefficients and displacement fields}
The  four interface continuity relations lead to
\begin{equation}\label{7-050}
\begin{array}{rcr}
  A^{[0]-}e^{[0]}-A^{[1]+}\big(e^{[1]}\big)^{-1}-A^{[1]-}e^{[1]} &=& -A^{[0]+}\big(e^{[0]}\big)^{-1} \\
  -A^{[0]-}e^{[0]}-A^{[1]+}G^{[10]}\big(e^{[1]}\big)^{-1}+A^{[1]-}G^{[10]}e^{[1]} &=& -A^{[0]+}\big(e^{[0]}\big)^{-1} \\
  A^{[1]+}+A^{[1]-}-A^{[2]+} &=& 0 \\
  A^{[1]+}-A^{[1]-}-A^{[2]+}G^{[21]} &=& 0
\end{array}
~,
\end{equation}
(with $e^{[j]}=\exp(ik_{z}^{[j]}h)$ and $G^{[jk]}=G^{[j]}/G^{[k]}$, $G^{[j]}=k_{z}^{[j]}/R^{[j]}$)
for the four unknowns $A^{[0]-},~A^{[1]+}$, $A^{[1]-},~A^{[2]+}$. The solution of (\ref{7-050})  is:
\begin{equation}\label{7-060}
A^{[0]-}=
-A^{[0]+}\big(e^{[0}\big)^{-2}\left[
\big(G^{[2]}+G^{[1]}\big)\big(G^{[1]}-G^{[0]}\big)\big(e^{[1]}\big)^{-1}+
\big(G^{[2]}-G^{[1]}\big)\big(G^{[1]}+G^{[0]}\big)e^{[1]}
\right]D^{-1}~,
\end{equation}
\begin{equation}\label{7-062}
A^{[1]+}=
A^{[0]+}\big(e^{[0}\big)^{-1}\left[
2G^{[0]}\big(G^{[1]}+G^{[2]}\big)
\right]D^{-1}~,
\end{equation}
\begin{equation}\label{7-064}
A^{[1]-}=
A^{[0]+}\big(e^{[0}\big)^{-1}\left[
2G^{[0]}\big(G^{[1]}-G^{[2]}\big)
\right]D^{-1}~,
\end{equation}
\begin{equation}\label{7-070}
A^{[2]+}=
A^{[0]+}\big(e^{[0}\big)^{-1}
\big[4G^{[1]}G^{[0]}\big]
D^{-1}~,
\end{equation}
wherein
\begin{equation}\label{7-080}
D=\big(G^{[2]}+G^{[1]}\big)\big(G^{[1]}+G^{[0]}\big)\big(e^{[1]}\big)^{-1}+
  \big(G^{[2]}-G^{[1]}\big)\big(G^{[1]}-G^{[0]}\big)e^{[1]}~.
\end{equation}
These expressions are easily cast into the following forms:
\begin{equation}\label{7-085}
A^{[0]-}=
-A^{[0]+}\big(e^{[0}\big)^{-2}\left[
\frac{G^{[1]}\big(\big(G^{[2]}-G^{[0]}\big)C+\big(G^{[2]}G^{[0]}-G^{[1]}G^{[1]}\big)iS}
     {G^{[1]}\big(\big(G^{[2]}+G^{[0]}\big)C-\big(G^{[2]}G^{[0]}+G^{[1]}G^{[1]}\big)iS}
\right]~,
\end{equation}
\begin{equation}\label{7-095}
A^{[1]+}=
A^{[0]+}\big(e^{[0}\big)^{-1}\left[
\frac{G^{[0]}\big(G^{[1]}+G^{[2]}\big)}
     {G^{[1]}\big(\big(G^{[2]}+G^{[0]}\big)C-\big(G^{[2]}G^{[0]}+G^{[1]}G^{[1]}\big)iS}
\right]~,
\end{equation}
\begin{equation}\label{7-105}
A^{[1]-}=
A^{[0]+}\big(e^{[0}\big)^{-1}\left[
\frac{G^{[0]}\big(G^{[1]}-G^{[2]}\big)}
     {G^{[1]}\big(\big(G^{[2]}+G^{[0]}\big)C-\big(G^{[2]}G^{[0]}+G^{[1]}G^{[1]}\big)iS}
\right]~,
\end{equation}
\begin{equation}\label{7-115}
A^{[2]+}=
A^{[0]+}\big(e^{[0}\big)^{-1}\left[
\frac{2G^{[2]}G^{[0]}}
     {G^{[1]}\big(\big(G^{[2]}+G^{[0]}\big)C-\big(G^{[2]}G^{[0]}+G^{[1]}G^{[1]}\big)iS}
\right]~.
\end{equation}
wherein $S=\sin(k_{z}^{[1]}h)$, $C=\cos(k_{z}^{[1]}h)$.
%
\subsection{Comparison of the layer response to the grating response}
We notice  that
\begin{equation}\label{7-100}
\big(e^{[0]}\big)^{\pm 1}=\varepsilon_{0}^{\pm}~~,
\end{equation}
because of the assumption made at the outset that $C^{[l]}=c^{[l]}~;~l=0,2$ which implies that
\begin{equation}\label{7-102}
k_{z}^{[l]}=k_{z0}^{[l]}~;~l=0,2~.
\end{equation}
If, in addition, we assume that
\begin{equation}\label{7-104}
C^{'[1]}=c^{'[1]}~,
\end{equation}
\begin{equation}\label{7-105}
C^{''[1]}=c^{''[1]}~,
\end{equation}
then
\begin{equation}\label{7-106}
k_{z}^{[1]}=k_{z0}^{[1]}~\Rightarrow S=S_{0}~~,~~C=C_{0}~.
\end{equation}
Finally, if we assume (with $\phi=w/d$ the so-called filling factor) that
\begin{equation}\label{7-108}
R^{[1]}=\rho^{[1]}\phi^{-1}~,
\end{equation}
and recall that we assumed at the outset that
\begin{equation}\label{7-110}
R^{[l]}=\rho^{[l]}~;~l=0,2,
\end{equation}
 then
\begin{equation}\label{7-115}
G^{[l]}=g_{l}~;~l=0,1,2~,
\end{equation}
so that the comparison of (\ref{7-085})-(\ref{7-115}) with (\ref{4-400}), (\ref{4-410}) and (\ref{4-480})
shows that
\begin{equation}\label{7-120}
A_{0}^{[0]-(1)}=A^{[0]-}~~,~~A_{0}^{[1]\pm (1)}=A^{[0]\pm}~~,~~A_{0}^{[2]\pm (1)}=A^{[2]\pm}
~.
\end{equation}
The conditions (\ref{7-104}) and (\ref{7-108}) are what  one would expect to obtain from a mixture theory \cite{si02} approach to homogenization.

 Also the comparison of (\ref{4-480})-(\ref{4-500}) with (\ref{7-017})-(\ref{7-020}) shows, on account also of \ref{6-090}), (\ref{7-102}) and (\ref{7-106}) that
\begin{equation}\label{7-130}
U^{[l]}=u^{[l](1)}~;~=0,1,2
~,
\end{equation}
which indicates equality of the first-order approximation of the fields in the grating  configuration with the corresponding fields in the homogeneous layer configuration when (\ref{7-104}) and (\ref{7-108}) prevail.

 This means that  the first-order iteration approximation amounts to replacing the transmission grating by a homogeneous layer. Furthermore, the mass  density of this 'homogenized' layer is simply the mass density of the generic grating block $\rho^{[1]}$ divided by the filling factor $\phi$, all other parameters of the layer (thickness $h$, bulk wave velocity $C^{[1]}$ of the layer, bulk wave velocities $C^{[0]}$ and $C^{[2]}$ of the bottom and top half spaces, mass densities $\rho^{[0]}$ and $\rho^{[2]}$ of the bottom and top half spaces, incident angle $\theta^{i}$, and amplitude spectrum $A^{[0]+}(\omega)$ of the acoustic solicitation, being the same as for the transmission grating configuration.
\subsection{Conservation principle for the layer}\label{Consv}
Again using Green's second identity, leads in rather straightforward manner to  the following conservation principle:
\begin{equation}\label{7-120}
R+A+T=1~,
\end{equation}
wherein $R$ and $T$ are the hemisperical= single-wave reflected and transmitted fluxes respectively,  given by:
\begin{equation}\label{7-130}
R=\frac{\|A^{[0]-}\|^{2}}{\|A^{[0]+}\|^{2}}~~,~~
T=\frac{\|A^{[2]-}\|^{2}k_{z}^{[2]}R^{[0]}}{\|A^{[0]+}\|^{2}k_{z}^{[0]}R^{[2]}}~,
\end{equation}
and $A$ the absorbed flux given by
\begin{equation}\label{7-135}
A=\frac{1}{k_{z}^{[0]}d}\frac{R^{[0]}}{R^{[1]}}
\Im\left[\left(K^{[1]}\right)^{2}\right]\int_{\Omega_{1d}}\left\|\frac{U^{[1]}(\mathbf{x},\omega)}{A^{[0]+}(\omega)}\right\|^{2}d\varpi~,
\end{equation}
with $\Omega_{1d}$ the portion of the layer situated between $x=-d/2$ and $x=d/2$.

Expression (\ref{7-130}) shows, unsurprisingly (since $k_{z}^{[0]}=k_{z}^{i}$ and $k_{z}^{[2]}$ are real) that the only contribution to $R$ stems from the single specularly-reflected  homogeneous plane wave(s) and the only contribution to $T$ stems from  the single transmitted homogeneous plane wave (there exist no other transmitted) wave(s)).  The angles of emergence of these observable homogeneous waves are $\theta^{-}=\arcsin(k_{x}/k^{[0]})=\theta^{i}$ and $\theta^{+}=\arcsin(k_{x}/k^{[2]})$ (note again that $\theta^{-}=\theta^{i}$ is the angle of specular reflection in the sense of Snell, and $\theta^{+}=\arcsin(k^{[0]}\sin\theta^{i}/k^{[2]})$ the angle of refraction in the sense of Fresnel).

Actually, (\ref{7-020}) can be resumed by the following conservation relation
\begin{equation}\label{5-055}
F_{out}=F_{in}~,
\end{equation}
wherein $F_{out}=R+A+T$ is the normalized output flux and $F_{in}=1$ the normalized input flux.oth to the rigorous and approximate (i.e., $l-$order iterate) solutions to the transmission grating problem.
\section{Numerical results}
The purpose of the numerical study is essentially to find out how well the surrogate layer model responses compare to the grating responses.
\subsection{Assumed parameters and general indications of the information contained in the graphs}
In  the following figures it is assumed, unless written otherwise, that:
$A^{i}=A^{[0]+}=1$, $\theta^{i}=0^{\circ},~d=0.004~m,~h=0.00475~m,~c^{[0]}=343~ms^{-1},~\rho^{[0]}=1.2~Kgm^{-3},~c^{'[1]}=500~ms^{-1},~c^{''[1]}=0~ms^{-1}~
\rho^{[1]}=2.4~Kgm^{-3},~c^{[2]}=343~ms^{-1},~\rho^{[2]}=1.2~Kgm^{-3},~C^{'[1]}=c^{'[1]},~C^{''[1]}=c^{''[1]},~R^{[1]}=\phi/\rho^{[1]}$. Recall that the sites and thicknesses of the grating and layer are identical as are their solicitations.

In all the figures, the blue curves are relative to the grating whereas the red curves correspond to the surrogate (homogeneous) layer.

In the (1,1) (left-hand, uppermost) panel the full curves depict the amplitude transfer functions $\|A_{0}^{[2]+}\|$ and $\|A^{[2]+}\|$  whereas the dashed curves depict $\|A_{0}^{[1]-}\|$ and $\|A^{[1]-}\|$.

In the (2,1) (left-hand, lowermost) panel the full curves depict $\|A_{0}^{[0]-}\|$ and $\|A^{[0]-}\|$  whereas the dashed curves depict $\|A_{0}^{[1]+}\|$ and $\|A^{[1]+}\|$.

In the (1,2) (middle, uppermost) panel the full curves depict the transmitted fluxes (=spectral transmittances) $\tau$ (blue) and $T$ (red).

In the (2,2) (middle, lowermost) panel the full curves depict the reflected fluxes (=spectral reflectances) $\rho$ (blue) and $R$ (red).

In the (1,3) (right-hand, uppermost) panel the full curves depict the input fluxes $f_{in}$ (blue), $F_{in}$ (red) whereas  dashed curves depict the output fluxes $f_{out}$ (blue), $F_{out} (red)$.
In the (2,3) (right-hand, lowermost) panel the full curves depict the absorbed fluxes (=spectral absorptances) $\alpha$ (blue) and $A$ (red).
\subsection{Response of transmission gratings with wide spaces between blocks}
\begin{figure}[ht]
\begin{center}
\includegraphics[width=12cm] {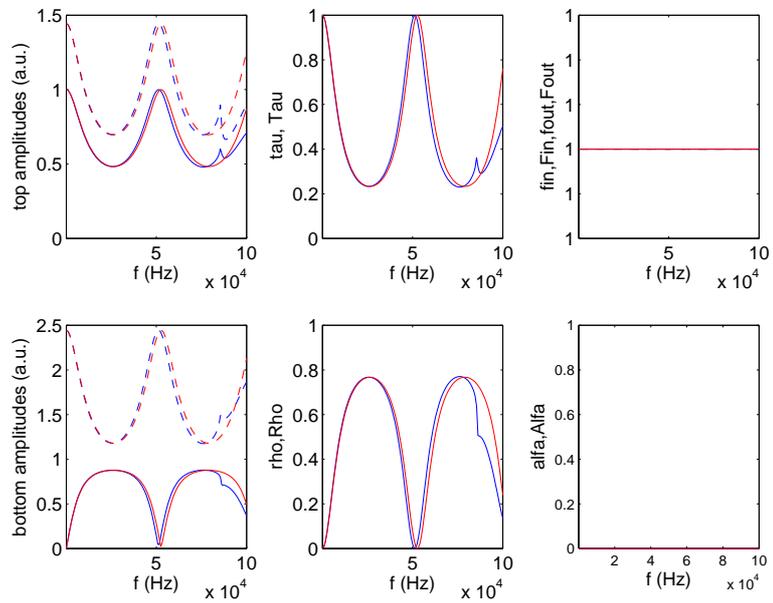}
 \caption{$C^{''[1]}=c^{''[1]}=0$}
  \label{cppwide-10}
  \end{center}
\end{figure}

\begin{figure}[ht]
\begin{center}
\includegraphics[width=12cm] {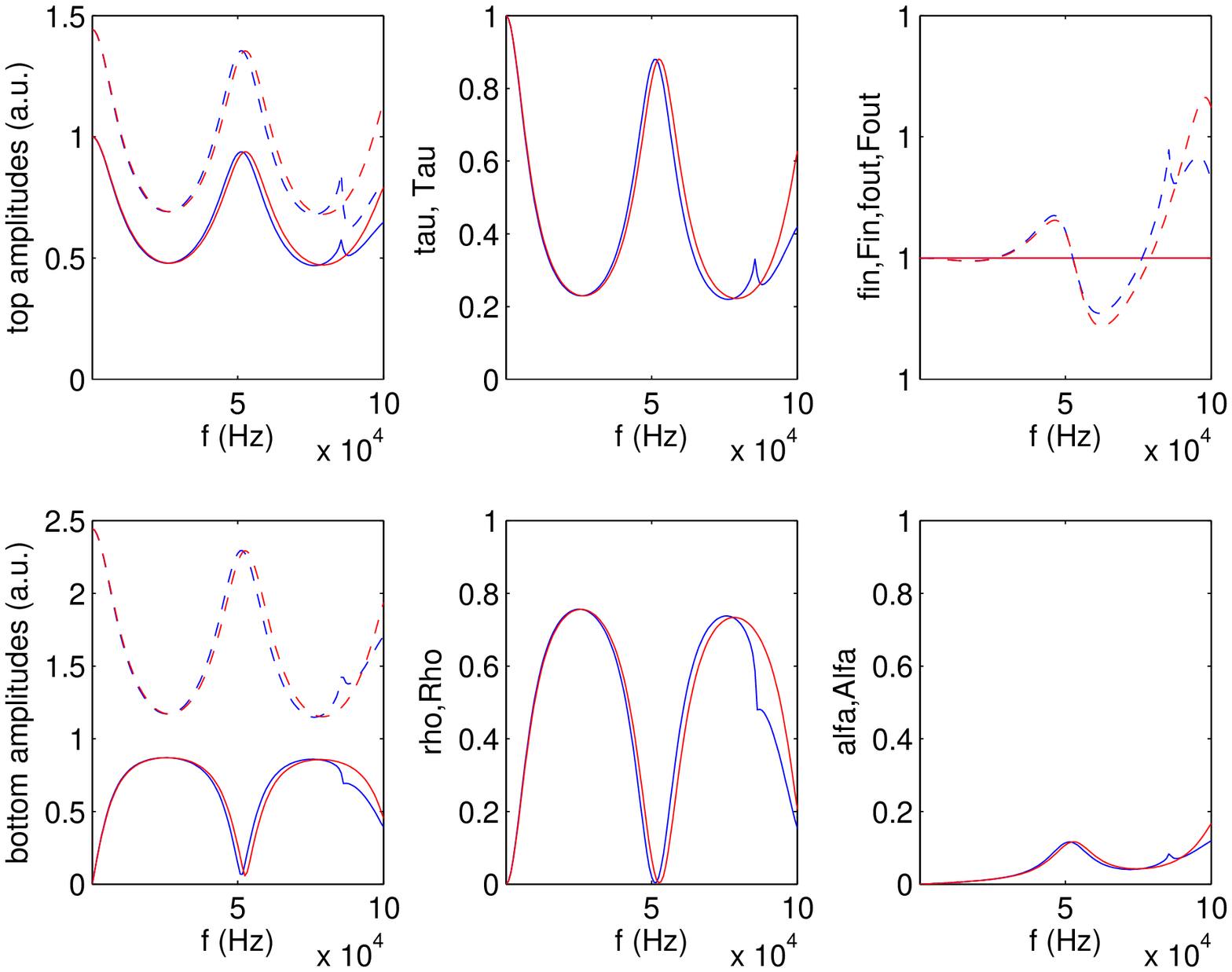}
 \caption{$C^{''[1]}=c^{''[1]}=-5$}
  \label{cppwide-20}
  \end{center}
\end{figure}

\begin{figure}[ht]
\begin{center}
\includegraphics[width=12cm] {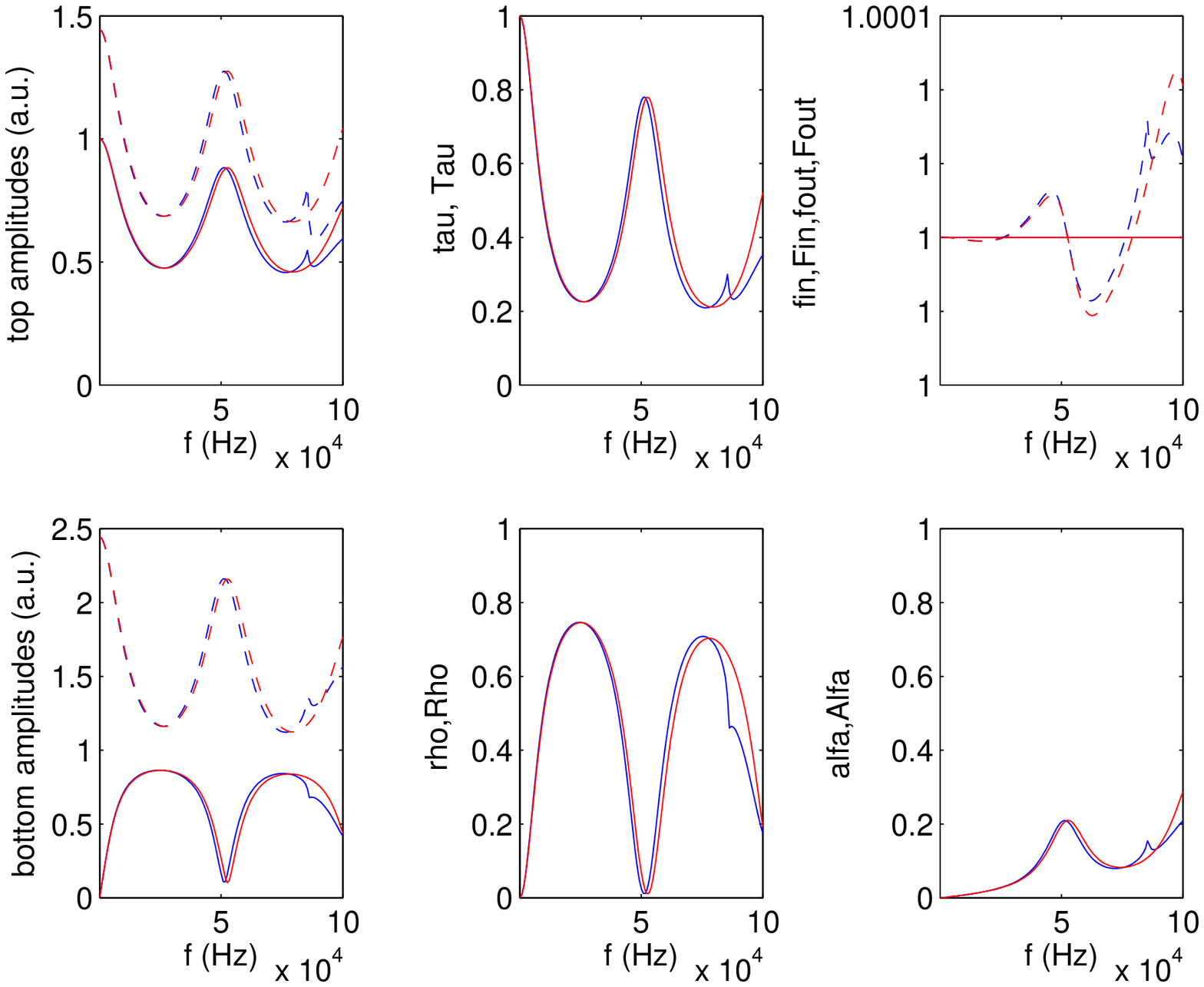}
 \caption{$C^{''[1]}=c^{''[1]}=-10$}
  \label{cppwide-30}
  \end{center}
\end{figure}

\begin{figure}[ht]
\begin{center}
\includegraphics[width=12cm] {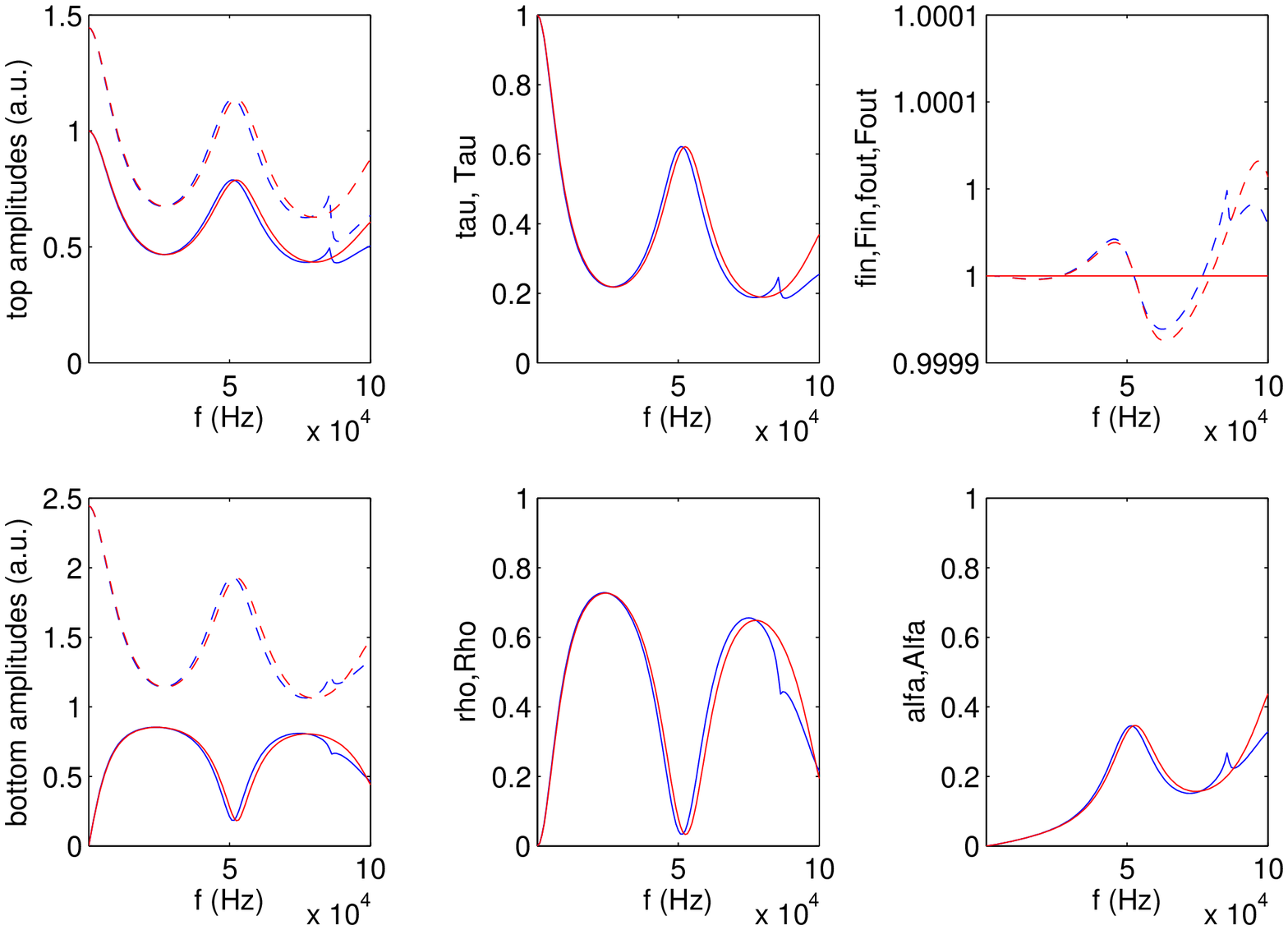}
 \caption{$C^{''[1]}=c^{''[1]}=-20$}
  \label{cppwide-40}
  \end{center}
\end{figure}

\begin{figure}[ht]
\begin{center}
\includegraphics[width=12cm] {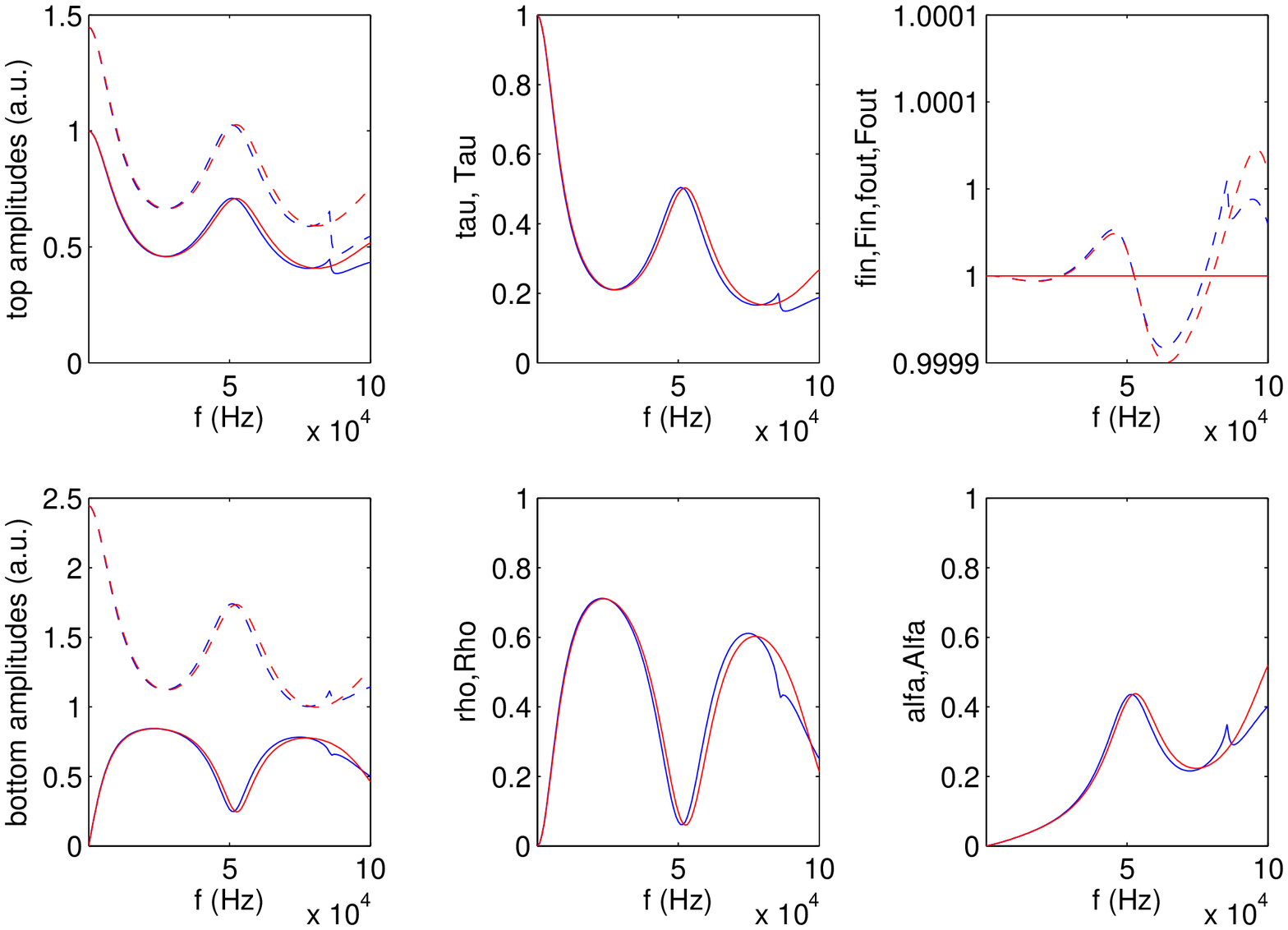}
 \caption{$C^{''[1]}=c^{''[1]}=-30$}
  \label{cppwide-50}
  \end{center}
\end{figure}

\begin{figure}[ht]
\begin{center}
\includegraphics[width=12cm] {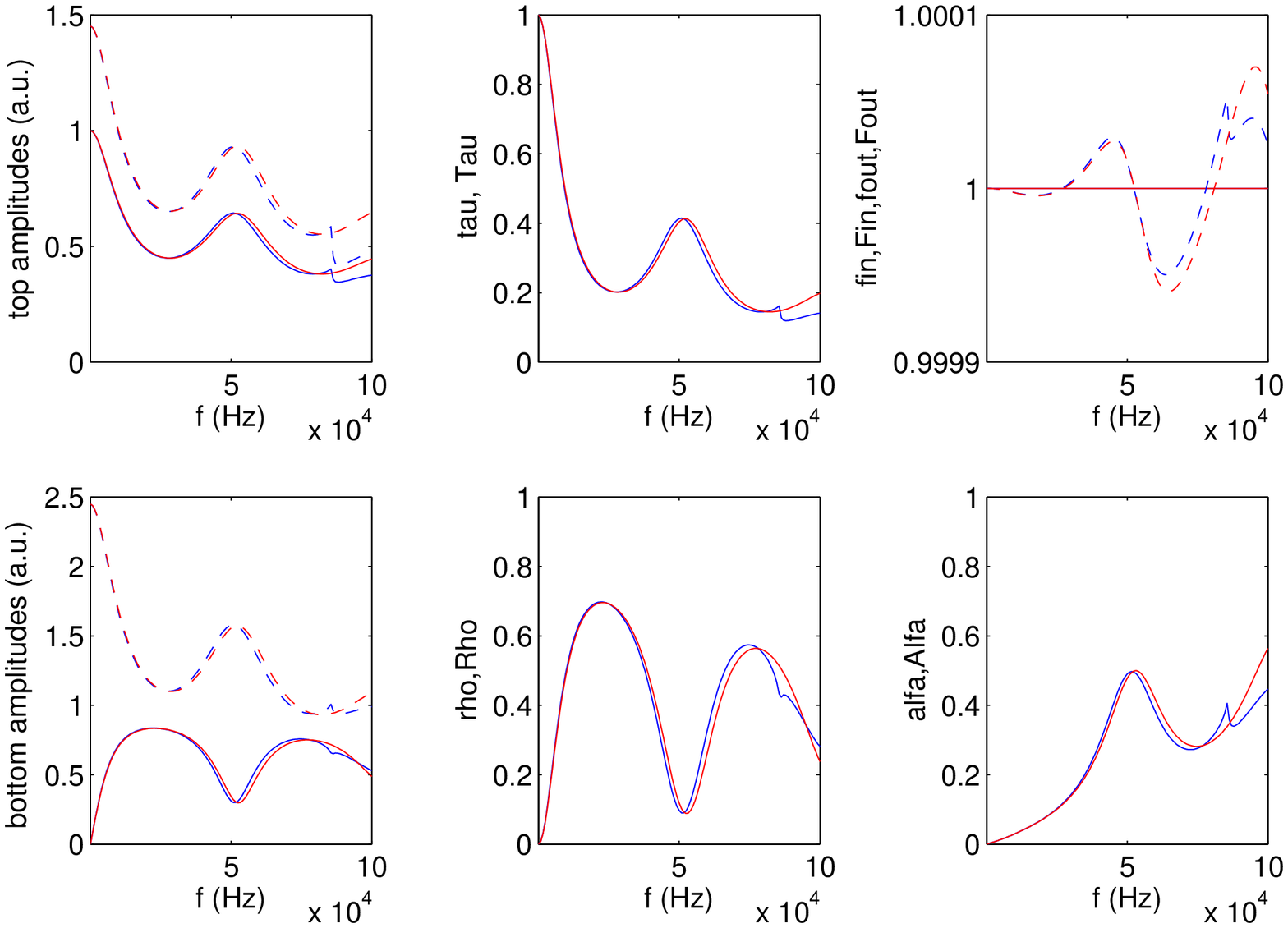}
 \caption{$C^{''[1]}=c^{''[1]}=-40$}
  \label{cppwide-60}
  \end{center}
\end{figure}

\begin{figure}[ht]
\begin{center}
\includegraphics[width=12cm] {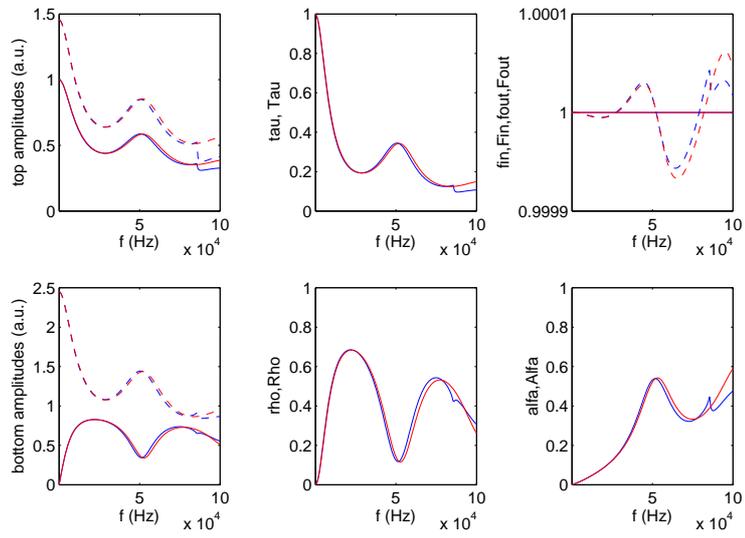}
 \caption{$C^{''[1]}=c^{''[1]}=-50$}
  \label{cppwide-70}
  \end{center}
\end{figure}

In figs. \ref{cppwide-10}-\ref{cppwide-70} we consider the case $w=0.003~m$ for various $C^{''[1]}=c^{''[1]}$.
\clearpage
\newpage
These figures show that the agreement between the grating (blue curves) and layer (red curves) responses is very good up till about $30 KHz$. This is as expected since the first-order iteration grating model= homogeneous layer model derives essentially from a low-frequency approximation. What is less expected is the rather good agreement between these two responses even beyond $30 KHz$ except in the neighborhood $f\approx 80.6 KHz$ of occurrence of a Wood anomaly \cite{ma16}. We note that flux is perfectly-well conserved for both configurations at all the considered frequencies.

Other noticeable features of the grating response, also present in the layer response, are: (i) the total transmission peak near $50 KHz$ when the interstitial material is lossless, (ii) the near-coincidence of frequencies of occurrence of the maxima of transmission and absorption,  (iii) the  nonlinear increase of absorption with the increase of $\|C^{''[1]}\|=\|c^{''[1]}\|$) and (iv) the fact that more than 35\% of the incident flux is absorbed beyond $f\approx 40~KHz$, with a peak of $55\%$ at $f\approx 55~KHz$, when $C^{''[1]}=c^{''[1]}=-50~ms^{-1}$.

\subsection{Response of transmission gratings with medium-width spaces between blocks}
\begin{figure}[ht]
\begin{center}
\includegraphics[width=12cm] {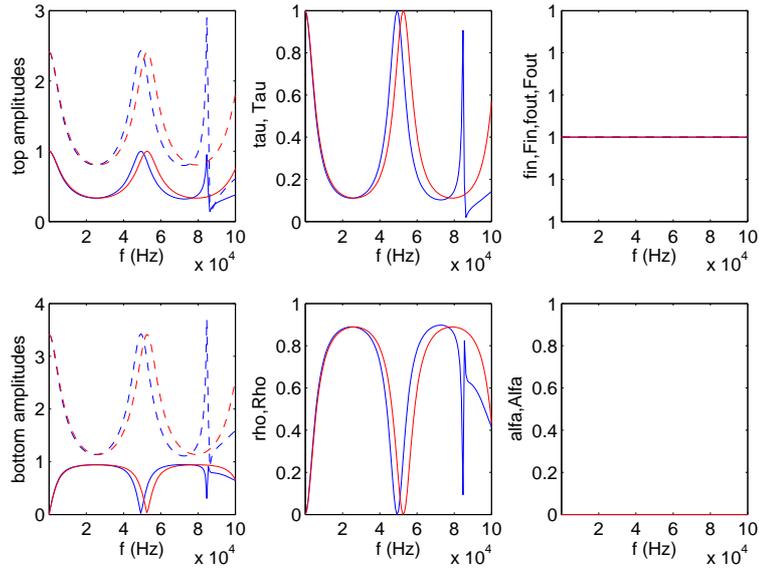}
 \caption{$C^{''[1]}=c^{''[1]}=0$}
  \label{cppmid-10}
  \end{center}
\end{figure}

\begin{figure}[ht]
\begin{center}
\includegraphics[width=12cm] {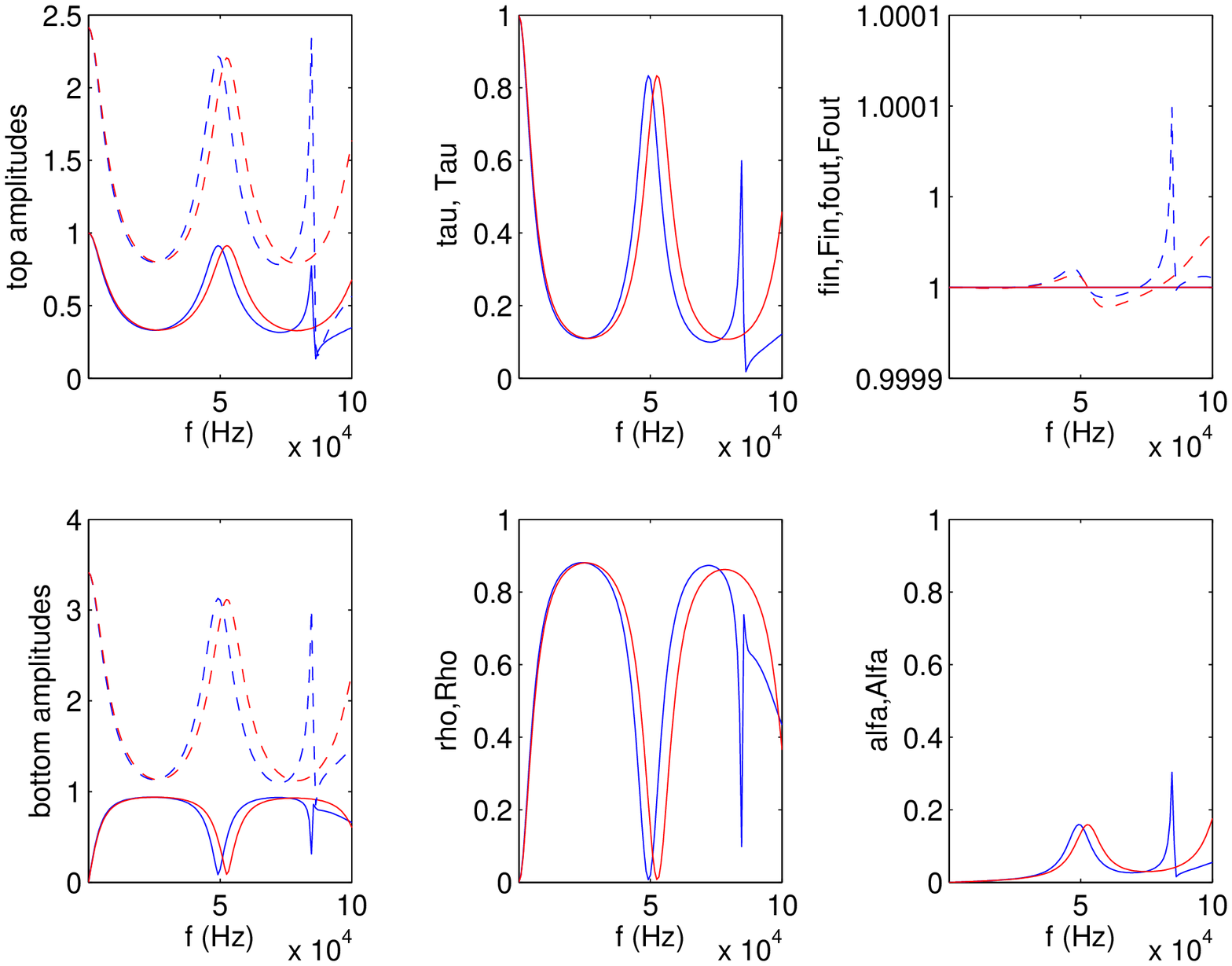}
 \caption{$C^{''[1]}=c^{''[1]}=-5$}
  \label{cppmid-20}
  \end{center}
\end{figure}

\begin{figure}[ht]
\begin{center}
\includegraphics[width=12cm] {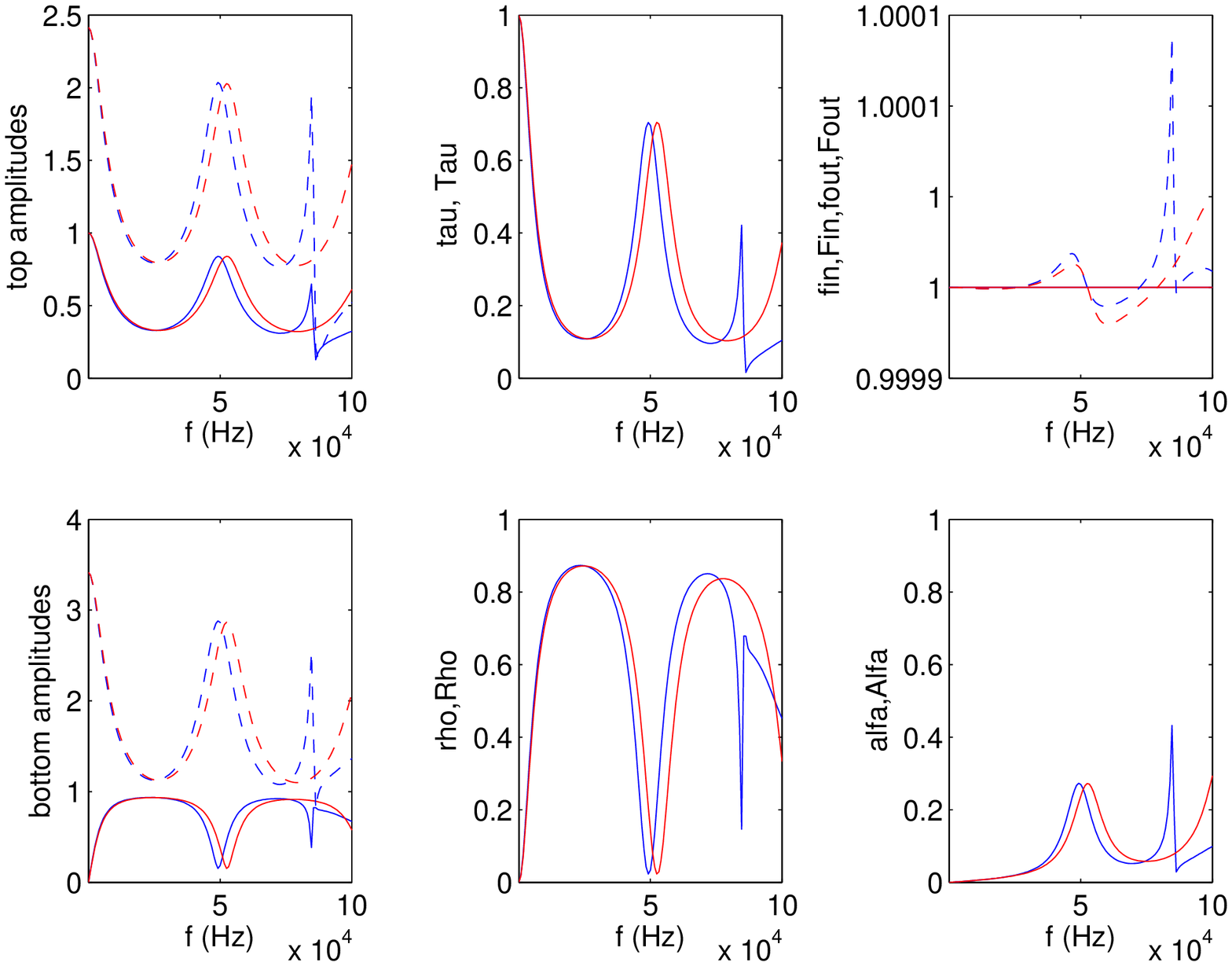}
 \caption{$C^{''[1]}=c^{''[1]}=-10$}
  \label{cppmid-30}
  \end{center}
\end{figure}

\begin{figure}[ht]
\begin{center}
\includegraphics[width=12cm] {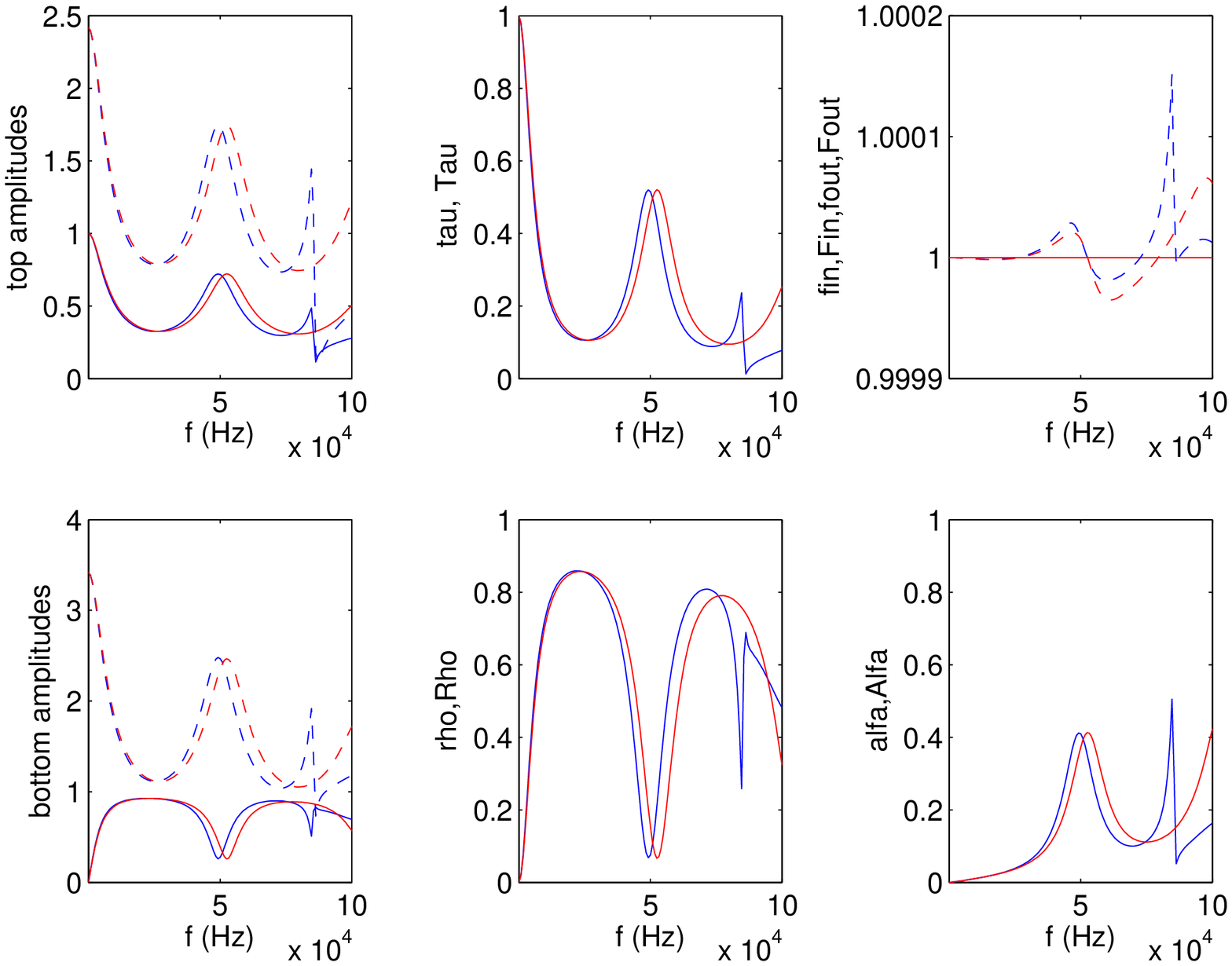}
 \caption{$C^{''[1]}=c^{''[1]}=-20$}
  \label{cppmid-40}
  \end{center}
\end{figure}

\begin{figure}[ht]
\begin{center}
\includegraphics[width=12cm] {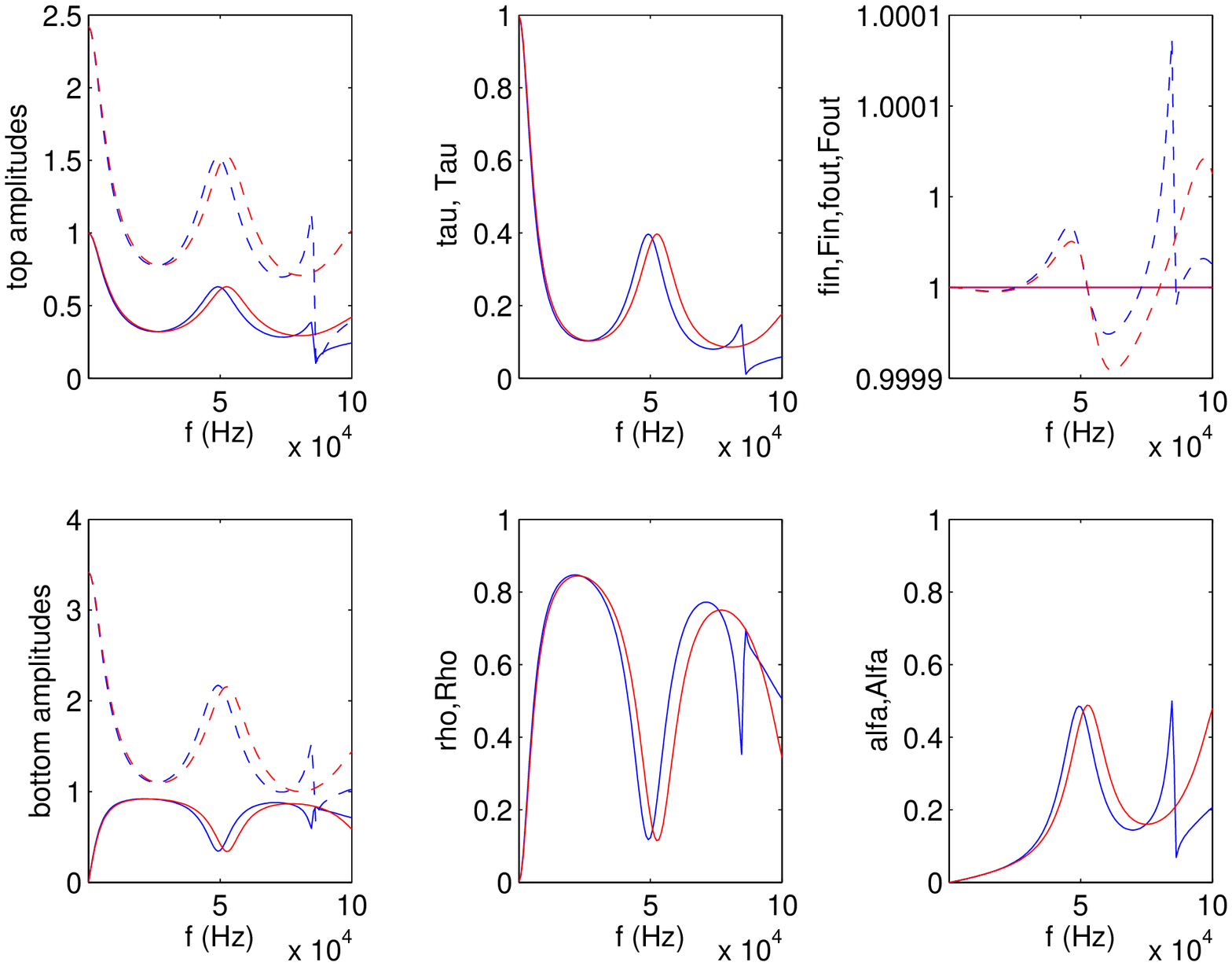}
 \caption{$C^{''[1]}=c^{''[1]}=-30$}
  \label{cppmid-50}
  \end{center}
\end{figure}

\begin{figure}[ht]
\begin{center}
\includegraphics[width=12cm] {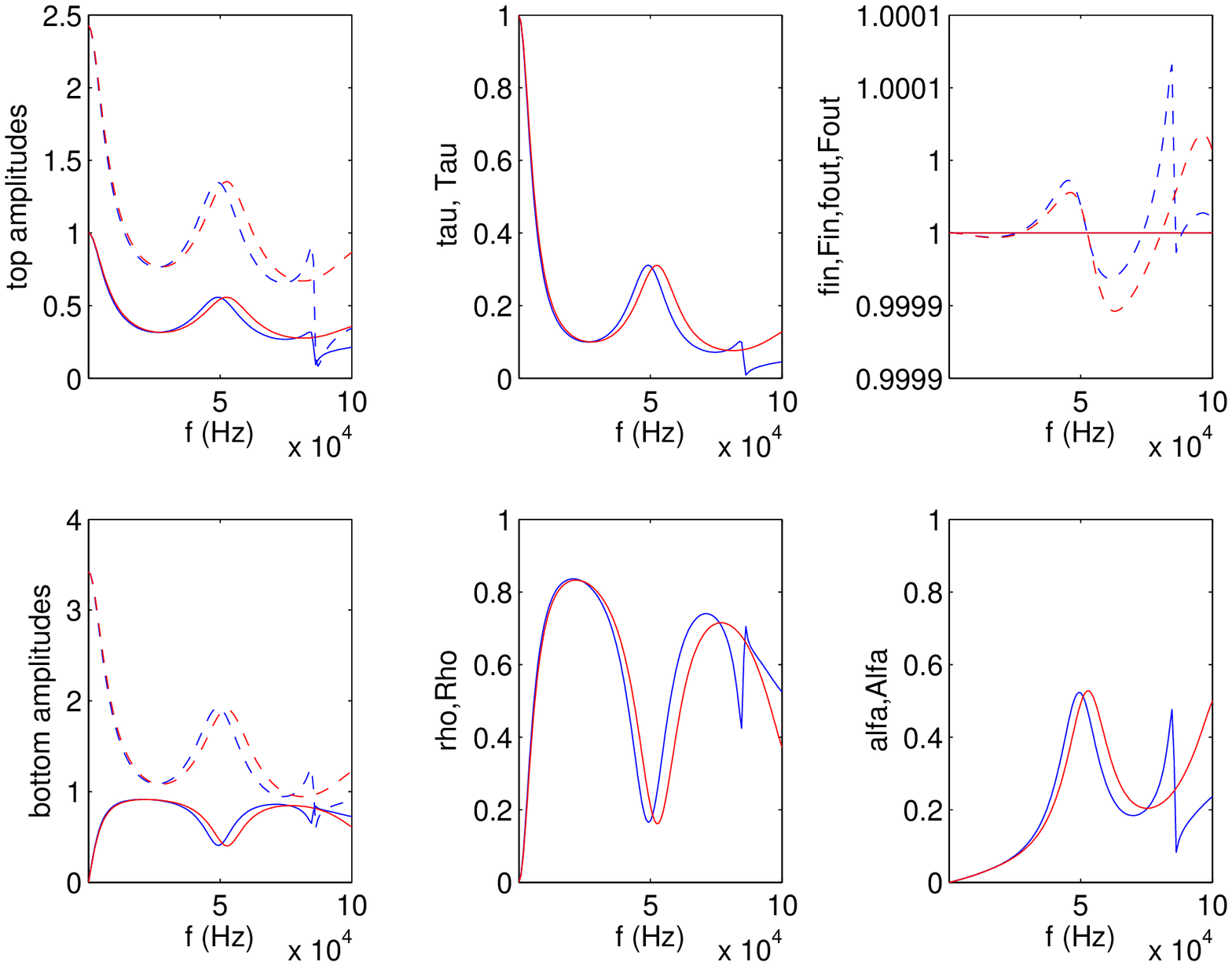}
 \caption{$C^{''[1]}=c^{''[1]}=-40$}
  \label{cppmid-60}
  \end{center}
\end{figure}

\begin{figure}[ht]
\begin{center}
\includegraphics[width=12cm] {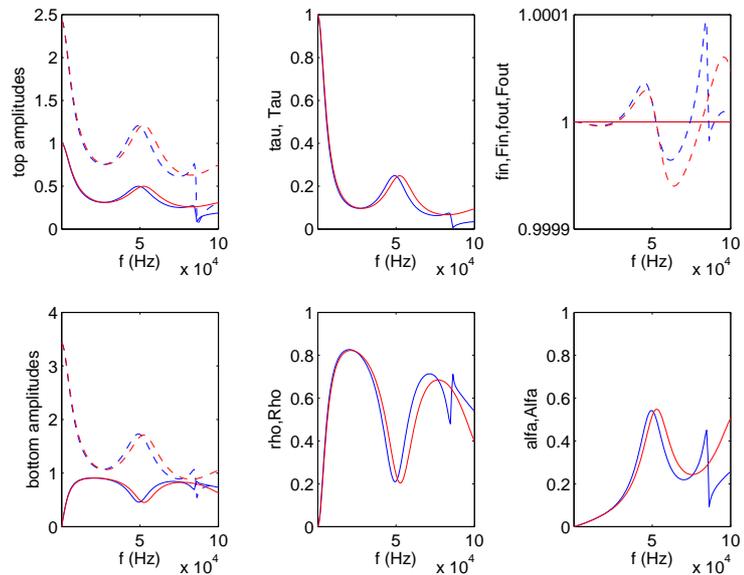}
 \caption{$C^{''[1]}=c^{''[1]}=-50$}
  \label{cppmid-70}
  \end{center}
\end{figure}

In figs. \ref{cppmid-10}-\ref{cppmid-70}  we consider the case $w=0.002~m$ for various $C^{''[1]}=c^{''[1]}$.
\clearpage
\newpage
These figures show that the agreement between the grating (blue curves) and layer (red curves) responses is very good up till about $25 KHz$. This is as expected since the first-order iteration grating model= homogeneous layer model derives essentially from a low-frequency approximation and is should fare less well for smaller $w/d$. What is less expected is the rather good agreement between these two responses even beyond $25 KHz$ except in the neighborhood $f\approx 80.6 KHz$ of occurrence of the Wood anomaly. We note that flux is nearly perfectly-well conserved for both configurations at all the considered frequencies.

Other noticeable features of the grating response, also present in the layer response, are: (i) the total transmission peak near $50 KHz$ when the interstitial material is lossless, (ii) the near-coincidence of frequencies of occurrence of the maxima of transmission and absorption,  (iii) the  nonlinear increase  of peak abosorption with the increase of $\|C^{''[1]}\|=\|c^{''[1]}\|$) and (iv) the fact that more than 10\% of the incident flux is absorbed beyond $f\approx 30~KHz$, with a peak of $55\%$ at $f\approx 53~KHz$, when $C^{''[1]}=c^{''[1]}=-50~ms^{-1}$.
\subsection{Response of transmission gratings with narrow spaces between blocks}
\begin{figure}[ht]
\begin{center}
\includegraphics[width=12cm] {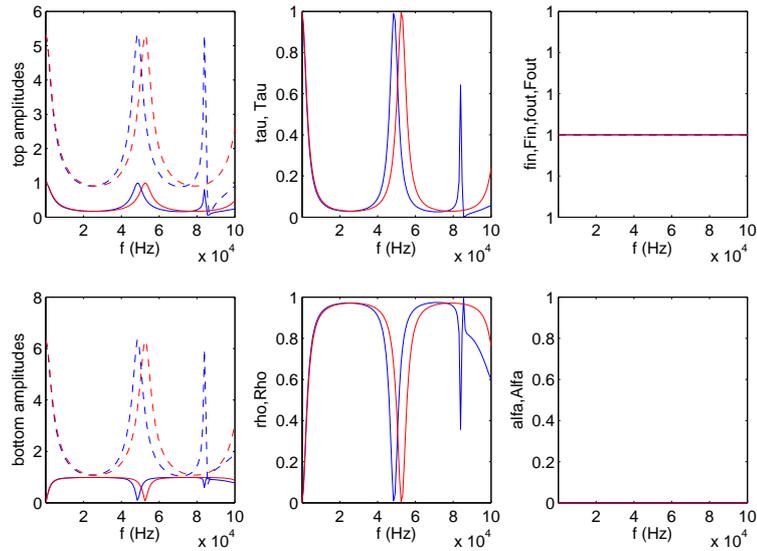}
 \caption{$C^{''[1]}=c^{''[1]}=0$}
  \label{cppnar-10}
  \end{center}
\end{figure}

\begin{figure}[ht]
\begin{center}
\includegraphics[width=12cm] {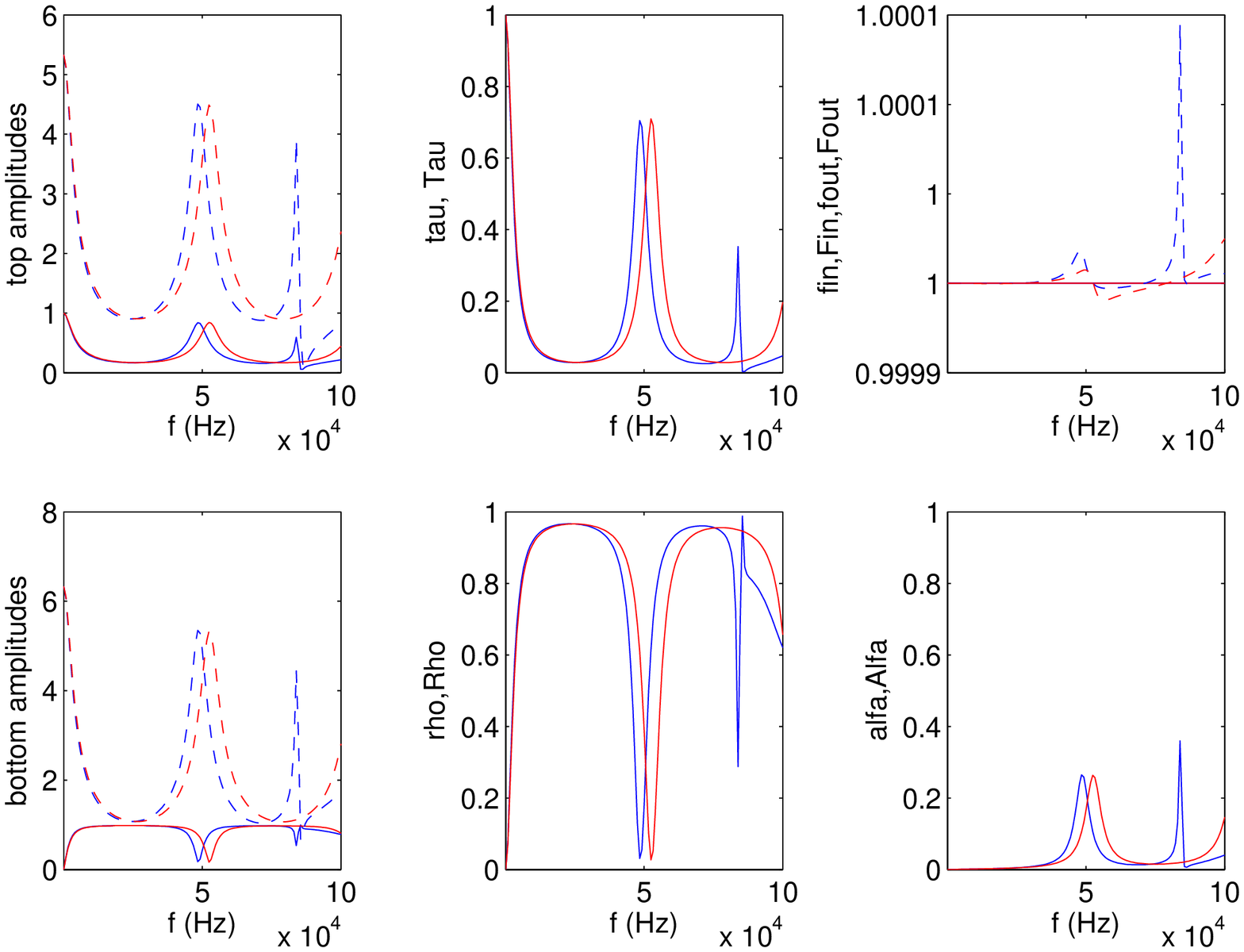}
 \caption{$C^{''[1]}=c^{''[1]}=-5$}
  \label{cppnar-20}
  \end{center}
\end{figure}

\begin{figure}[ht]
\begin{center}
\includegraphics[width=12cm] {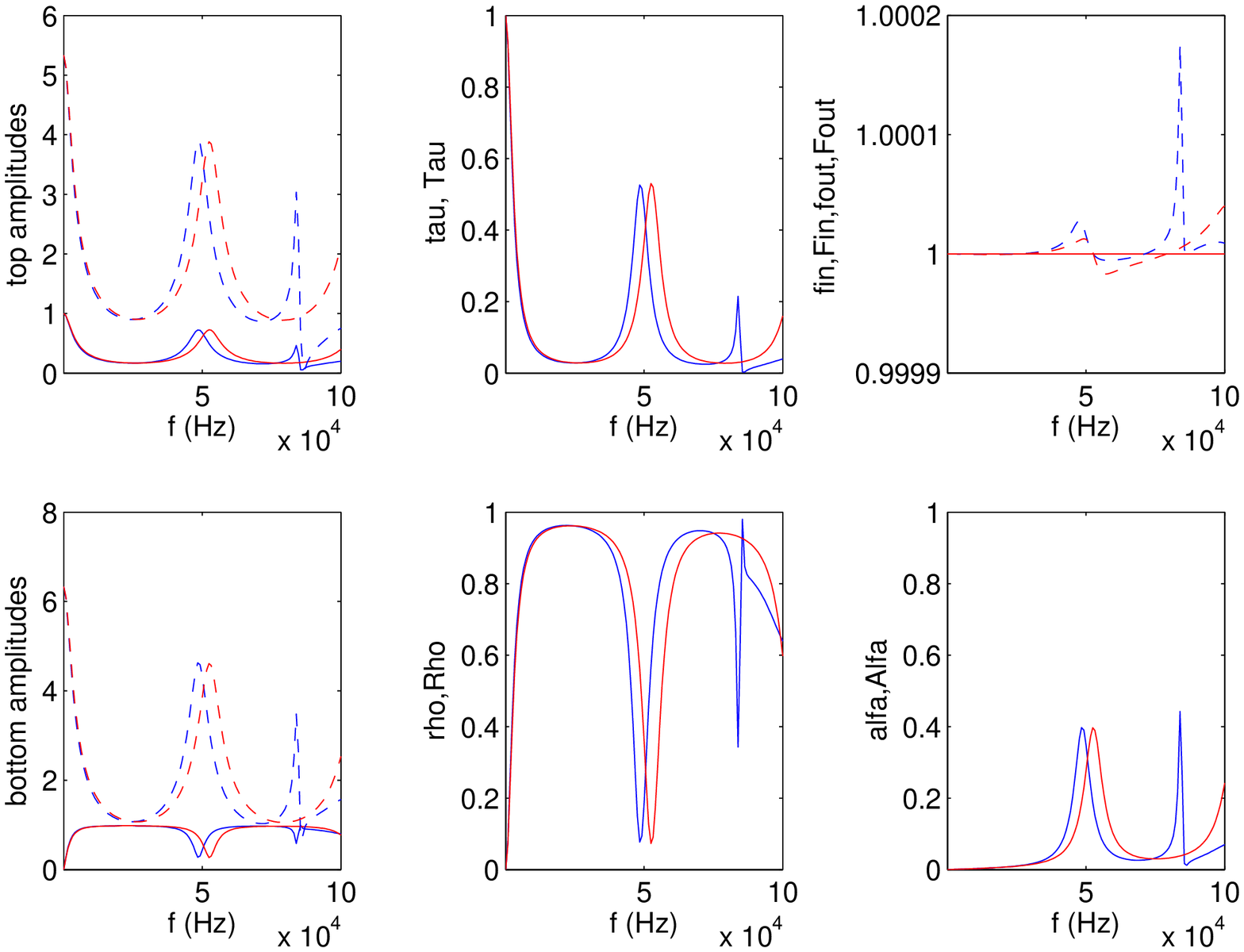}
 \caption{$C^{''[1]}=c^{''[1]}=-10$}
  \label{cppnar-30}
  \end{center}
\end{figure}

\begin{figure}[ht]
\begin{center}
\includegraphics[width=12cm] {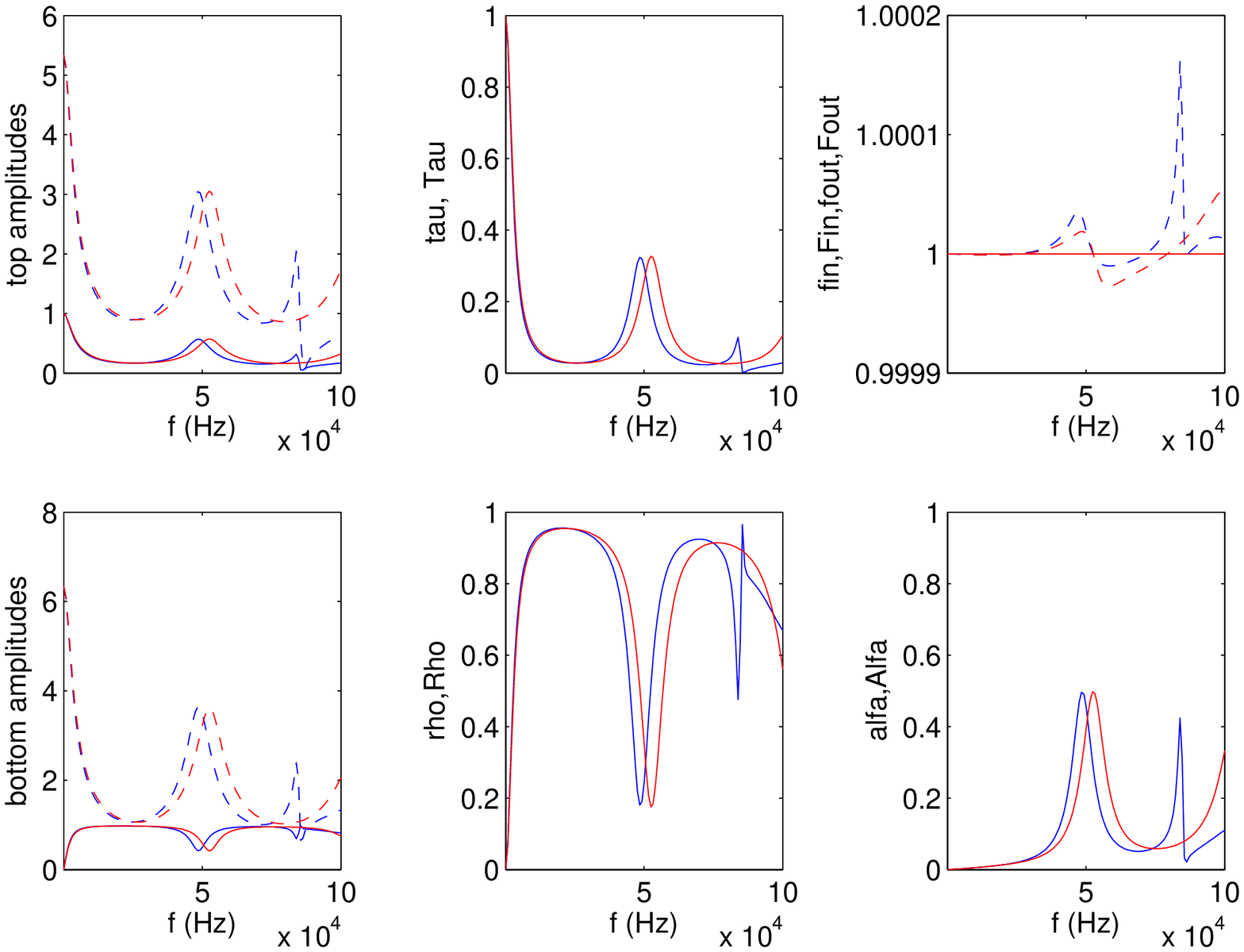}
 \caption{$C^{''[1]}=c^{''[1]}=-20$}
  \label{cppnar-40}
  \end{center}
\end{figure}

\begin{figure}[ht]
\begin{center}
\includegraphics[width=12cm] {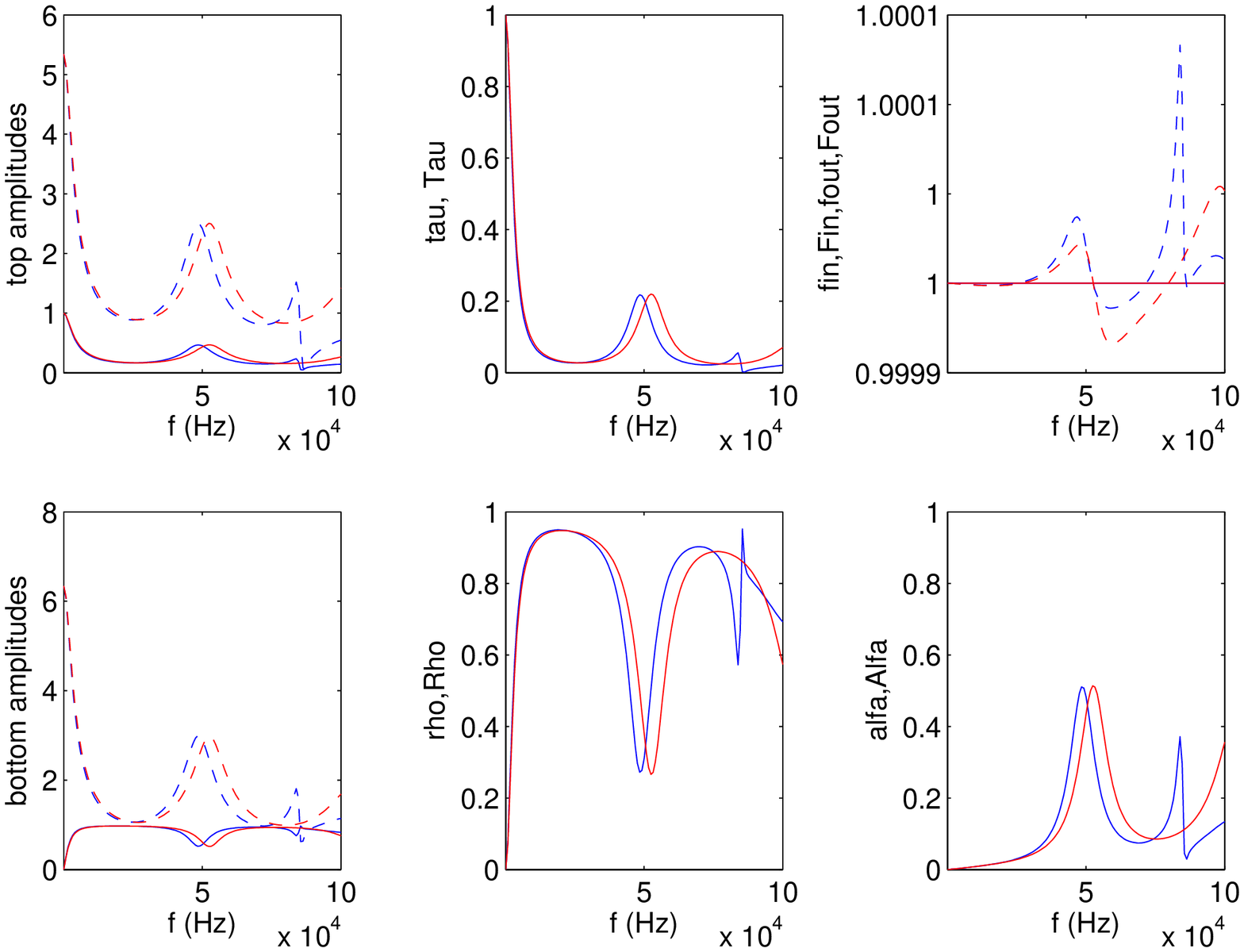}
 \caption{$C^{''[1]}=c^{''[1]}=-30$}
  \label{cppnar-50}
  \end{center}
\end{figure}

\begin{figure}[ht]
\begin{center}
\includegraphics[width=12cm] {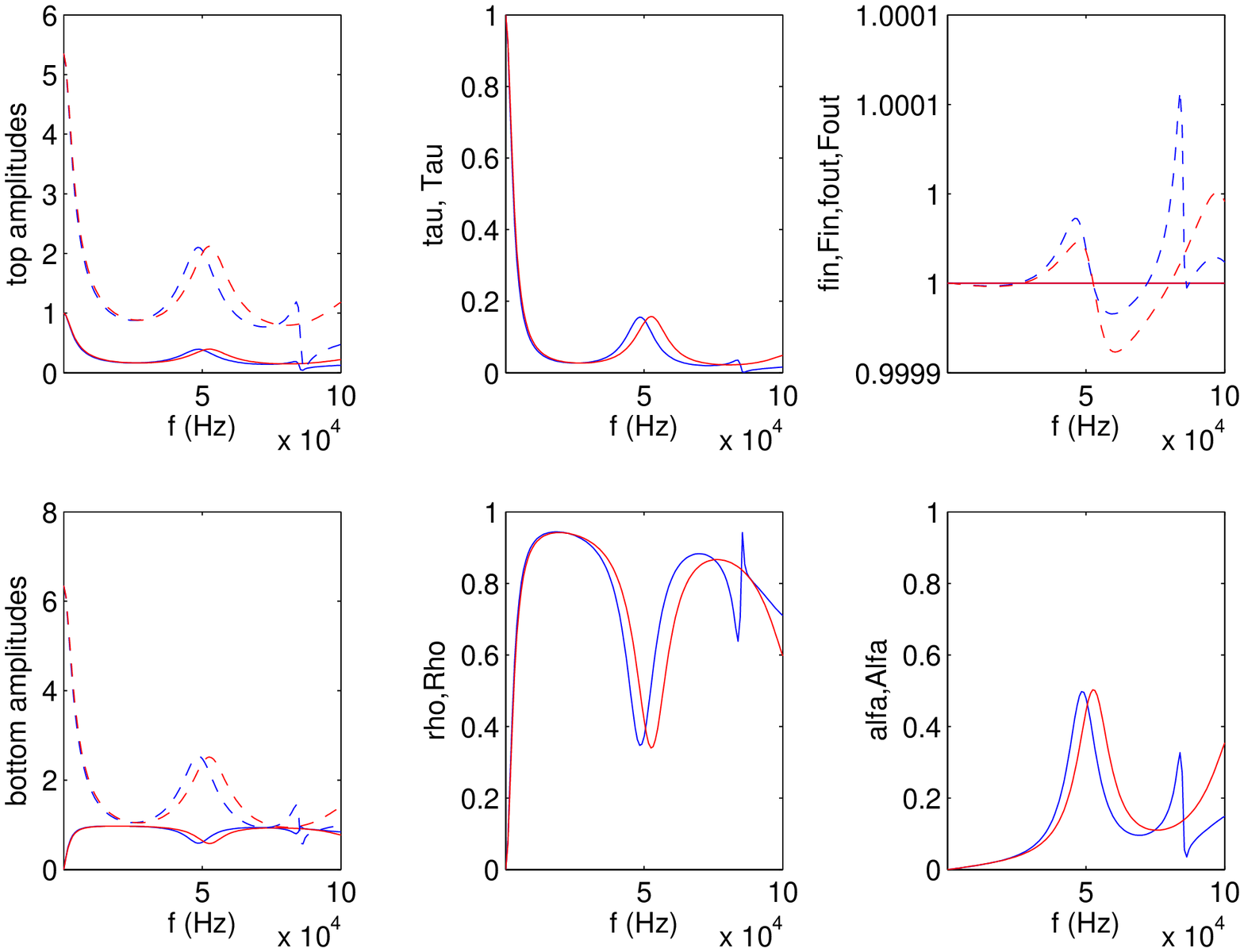}
 \caption{$C^{''[1]}=c^{''[1]}=-40$}
  \label{cppnar-60}
  \end{center}
\end{figure}

\begin{figure}[ht]
\begin{center}
\includegraphics[width=12cm] {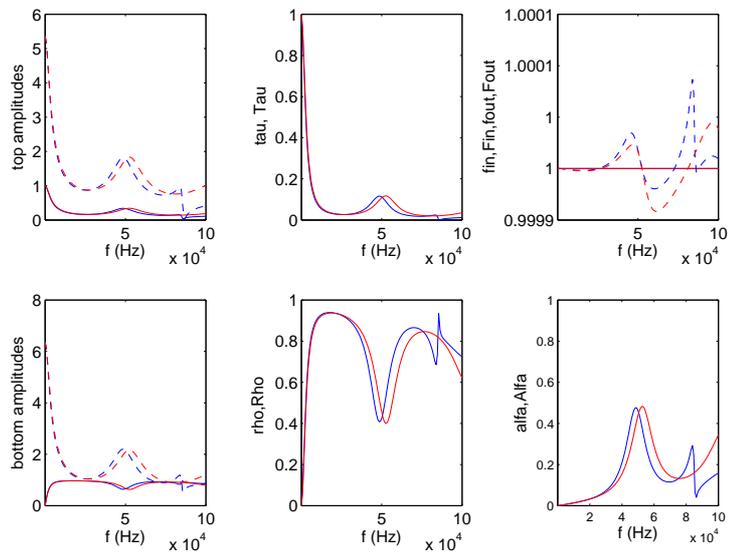}
 \caption{$C^{''[1]}=c^{''[1]}=-50$}
  \label{cppnar-70}
  \end{center}
\end{figure}

In figs. \ref{cppnar-10}-\ref{cppnar-70}  we consider the case $w=0.001~m$ for various $C^{''[1]}=c^{''[1]}$.
\clearpage
\newpage
These figures show that the agreement between the grating (blue curves) and layer (red curves) responses is very good up till about $20 KHz$. This is as expected since the first-order iteration grating model= homogeneous layer model derives essentially from a low-frequency approximation and is should fare less well for smaller $w/d$. What is less expected is the rather good agreement between these two responses even beyond $20 KHz$ except in the neighborhood $f\approx 80.6 KHz$ of occurrence of the Wood anomaly. We note that flux is  near perfectly-well conserved for both configurations at all the considered frequencies.

Other noticeable features of the grating response, also present in the layer response, are: (i) the total transmission peak near $50 KHz$ when the interstitial material is lossless, a feature that is unexpected for a grating with such narrow interstices, but in agreement with what has been predicted previously \cite{ll07} by finite element computations and verified experimentally, (ii) the near-coincidence of frequencies of occurrence of the maxima of transmission and absorption,  (iii) the  nonlinear increase, followed by leveling-off, of absorption with the increase of $\|C^{''[1]}\|=\|c^{''[1]}\|$) and (iv) the rather unexpected fact (considering the narrowness of the interstices between grating blocks) that more than 5\% of the incident flux is absorbed beyond $f\approx 20~KHz$, with a peak of $50\%$ at $f\approx 50~KHz$, when $C^{''[1]}=c^{''[1]}=-50~ms^{-1}$. The large absorption  is probably due to the existence of a very strong acoustic field within the interstices of the grating (and  therefore throughout the surrogate layer).
\subsection{Response of deep transmission gratings as a function of incident angle}
\begin{figure}[ht]
\begin{center}
\includegraphics[width=12cm] {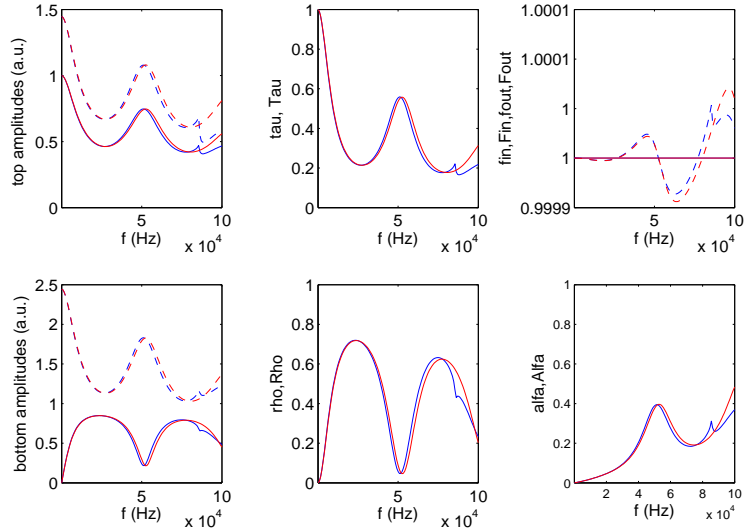}
 \caption{$\theta^{i}=0^{^{\circ}}$}
  \label{thideep-10}
  \end{center}
\end{figure}

\begin{figure}[ht]
\begin{center}
\includegraphics[width=12cm] {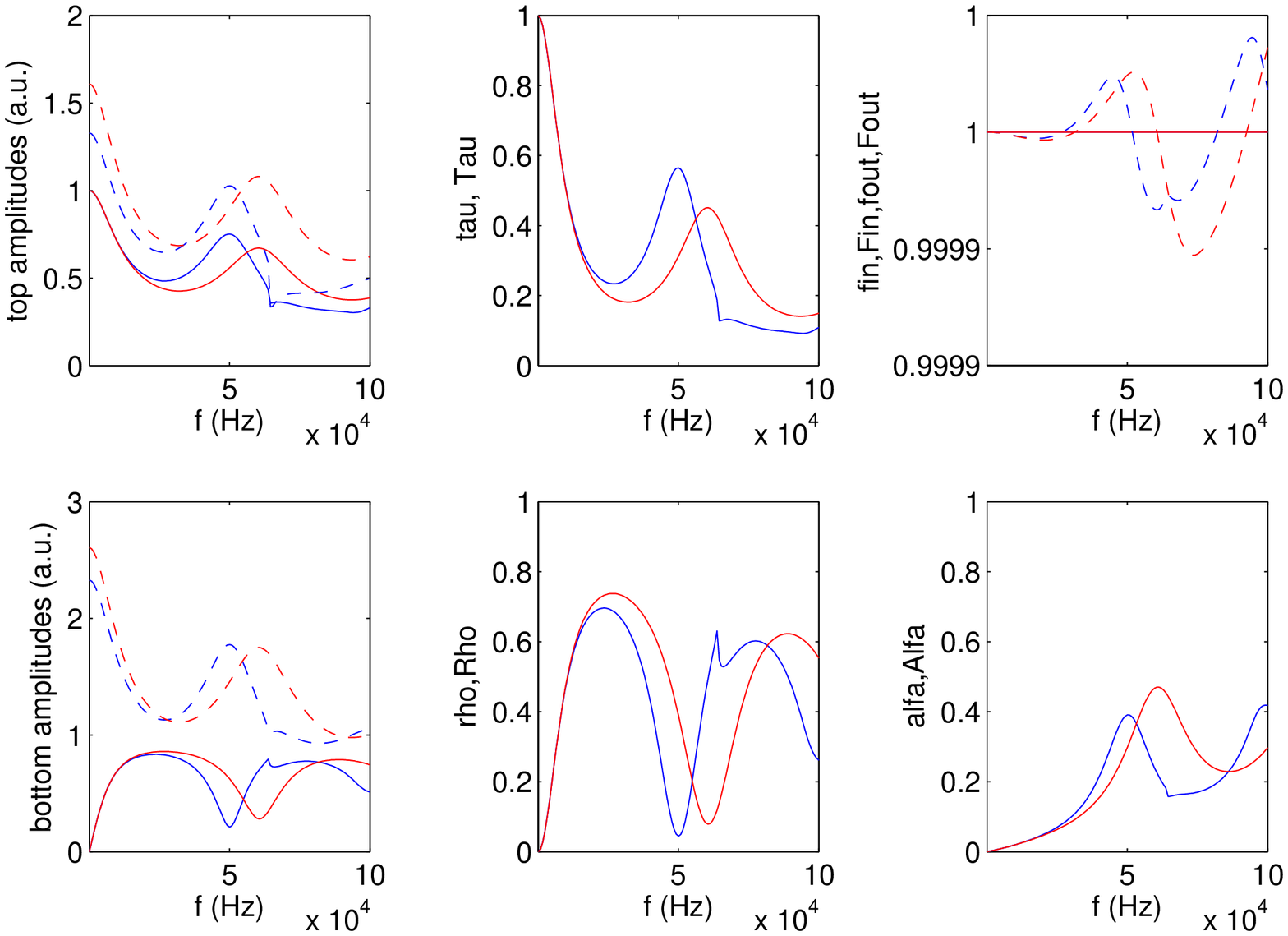}
 \caption{$\theta^{i}=20^{^{\circ}}$}
  \label{thideep-20}
  \end{center}
\end{figure}

\begin{figure}[ht]
\begin{center}
\includegraphics[width=12cm] {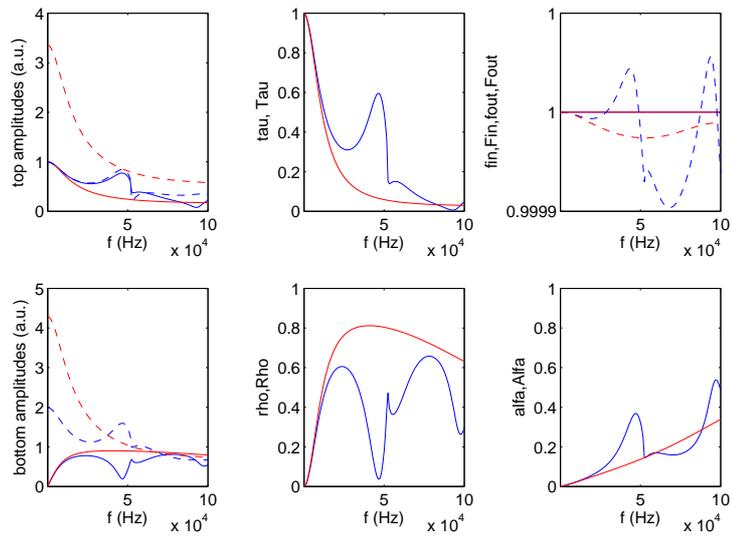}
 \caption{$\theta^{i}=40^{^{\circ}}$}
  \label{thideep-30}
  \end{center}
\end{figure}

In figs. \ref{thideep-10}-\ref{thideep-30}  we consider the case $w=0.003~m$, $h=0.00475~m$, $C^{''[1]}=c^{''[1]}=-25~ms^{-1}$ for various $\theta^{i}$.
\clearpage
\newpage
These figures show that the agreement between the two responses diminishes with increasing incident angle, notably because the frequency at which the lowest-order inhomogeneous waves in the half-spaces become homogeneous (this occurring at the frequencies of the Wood anomalies \cite{ma16}) diminishes with increasing $\theta^{i}$, so that the agreement is good up till $50~KHz$ for $\theta^{i}=0^{\circ}$, up till $\approx 15~KHz$ for $\theta^{i}=20^{\circ}$, and only up till $\approx 10~KHz$ for $\theta^{i}=40^{\circ}$. Nevertheless, there appears to exist agreement as to the secular trends of response (notably as concerns absorbed flux) between the two configurations for the three angles of incidence.

A last observation concerns the near-perfect conservation of flux for both the grating and equivalent layer at the three angles of incidence.
\subsection{Response of shallow transmission gratings as a function of the incident angle}
\begin{figure}[ht]
\begin{center}
\includegraphics[width=12cm] {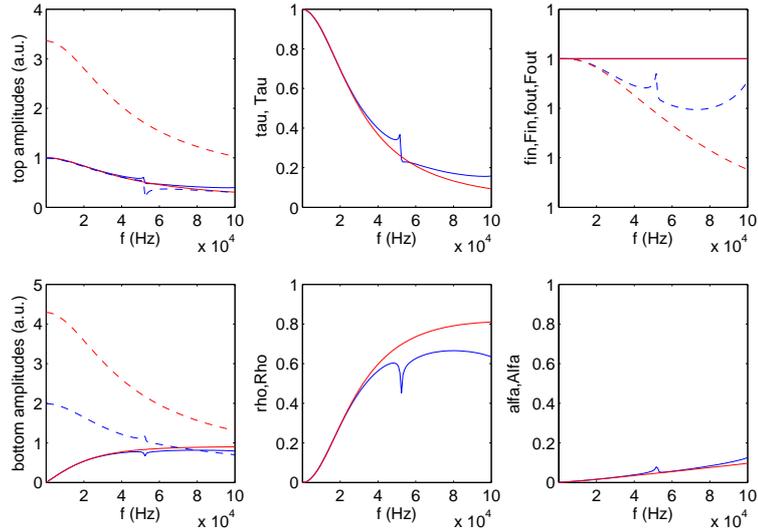}
 \caption{$\theta^{i}=0^{^{\circ}}$}
  \label{thishal-10}
  \end{center}
\end{figure}

\begin{figure}[ht]
\begin{center}
\includegraphics[width=12cm] {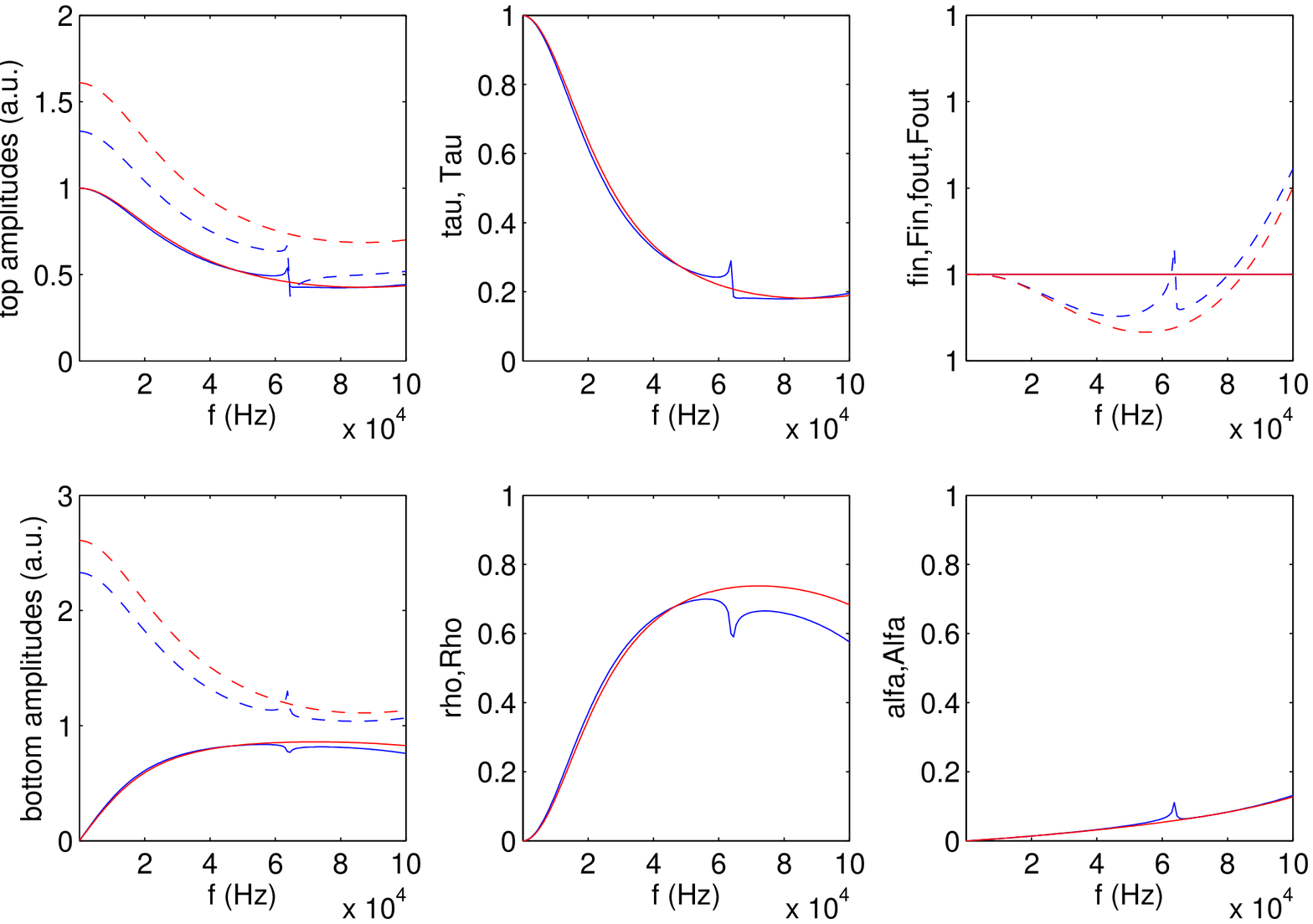}
 \caption{$\theta^{i}=20^{^{\circ}}$}
  \label{thishal-20}
  \end{center}
\end{figure}

\begin{figure}[ht]
\begin{center}
\includegraphics[width=12cm] {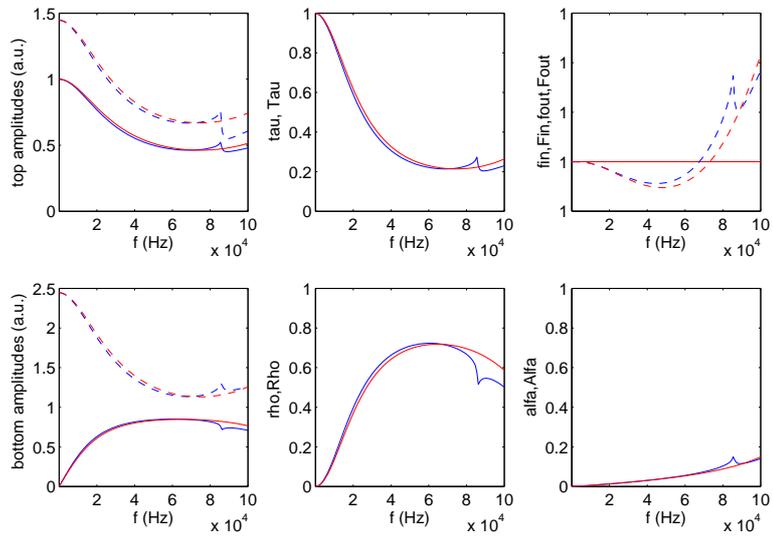}
 \caption{$\theta^{i}=40^{^{\circ}}$}
  \label{thishal-30}
  \end{center}
\end{figure}

In figs. \ref{thishal-10}-\ref{thishal-30}  we consider the case $w=0.003~m$, $h=0.00175~m$, $C^{''[1]}=c^{''[1]}=-25~ms^{-1}$ for various $\theta^{i}$.
\clearpage
\newpage
These figures show that, contrary to the case of deep blocks/layers, the agreement between the two responses is excellent for all frequencies up to the one at which the Wood anomaly occurs, this frequency diminishing with increasing angle of incidence. Again, there appears to exist agreement as to the secular trends of response (notably as concerns absorbed flux) between the two configurations for the three angles of incidence.

A last observation concerns the perfect conservation of flux for both the grating and equivalent layer at the three angles of incidence.
\subsection{Response of  transmission gratings as a function of their thickness}
\begin{figure}[ht]
\begin{center}
\includegraphics[width=12cm] {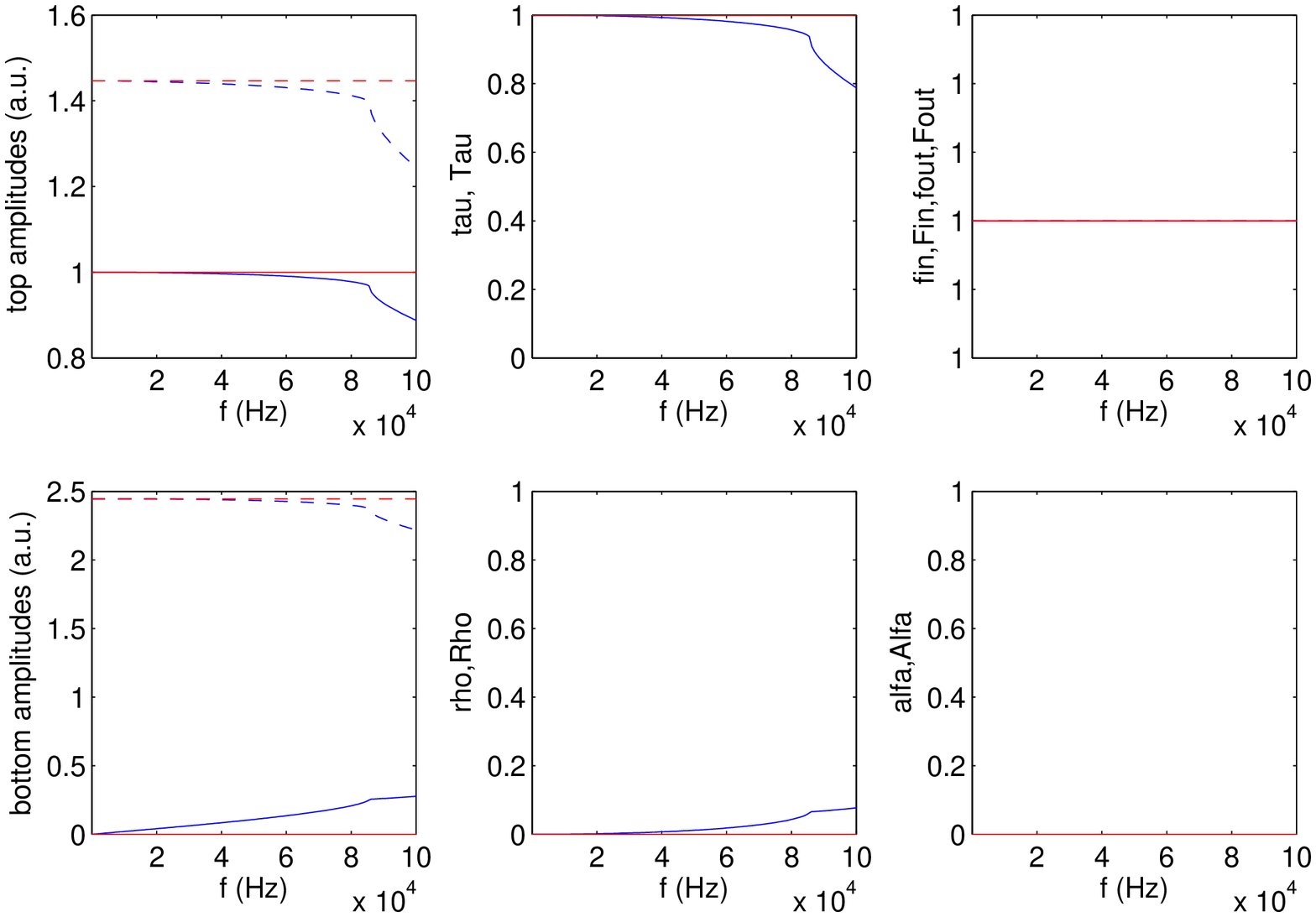}
 \caption{$h=0 m$}
  \label{h-10}
  \end{center}
\end{figure}

\begin{figure}[ht]
\begin{center}
\includegraphics[width=12cm] {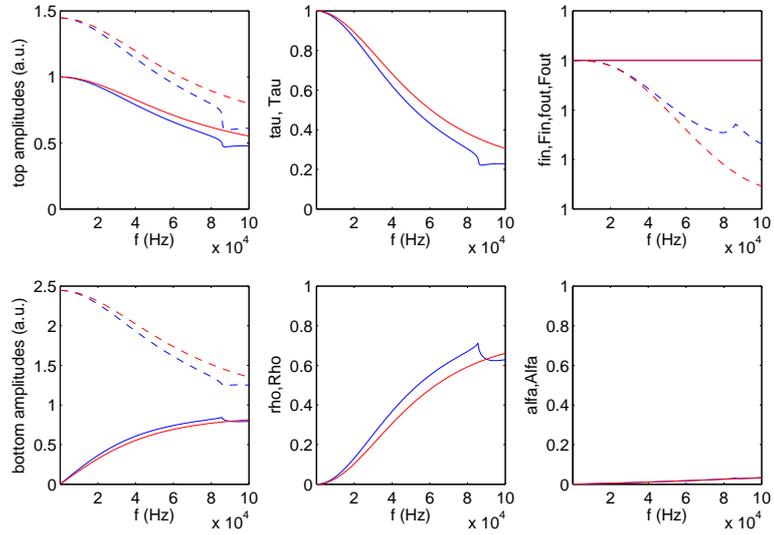}
 \caption{$h=0.00075 m$}
  \label{h-20}
  \end{center}
\end{figure}

\begin{figure}[ht]
\begin{center}
\includegraphics[width=12cm] {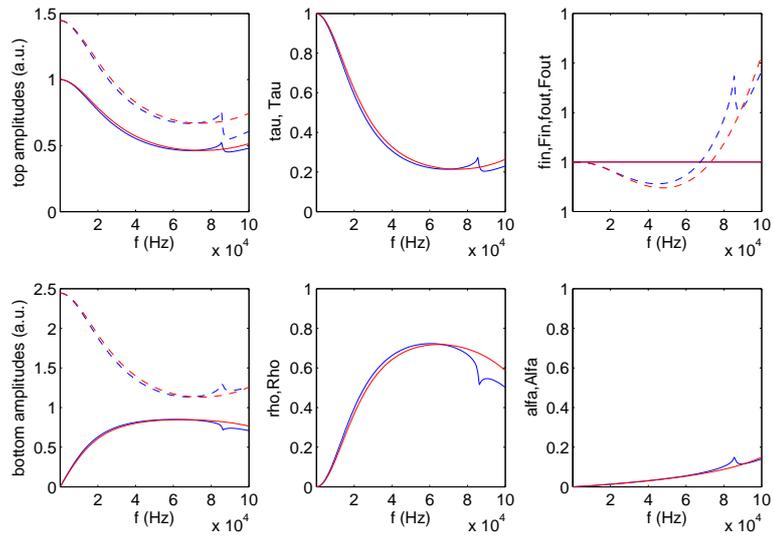}
 \caption{$h=0.00175 m$}
  \label{h-30}
  \end{center}
\end{figure}

\begin{figure}[ht]
\begin{center}
\includegraphics[width=12cm] {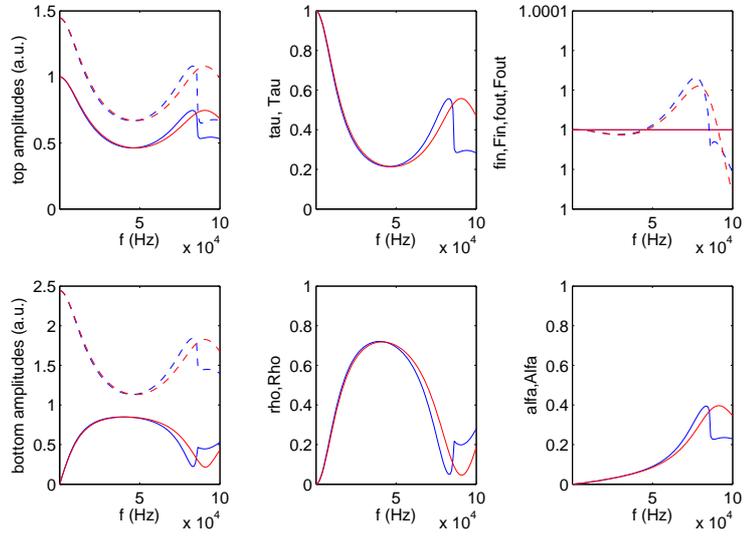}
 \caption{$h=0.00275 m$}
  \label{h-40}
  \end{center}
\end{figure}

\begin{figure}[ht]
\begin{center}
\includegraphics[width=12cm] {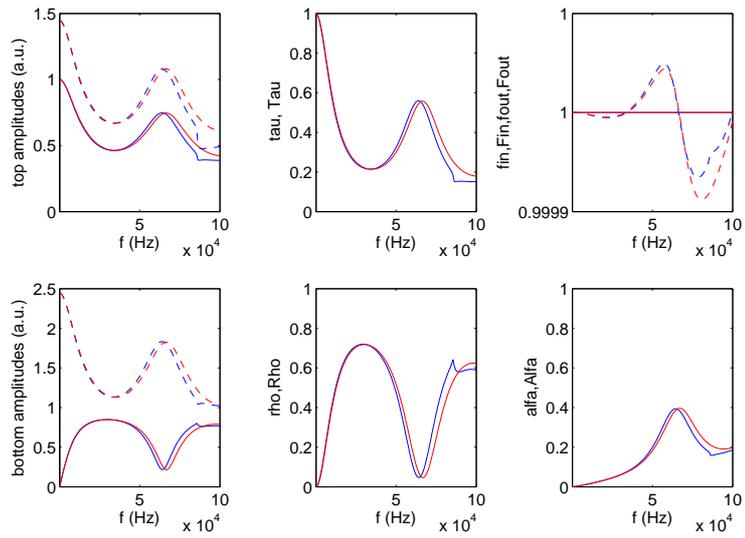}
 \caption{$h=0.00375 m$}
  \label{h-50}
  \end{center}
\end{figure}

\begin{figure}[ht]
\begin{center}
\includegraphics[width=12cm] {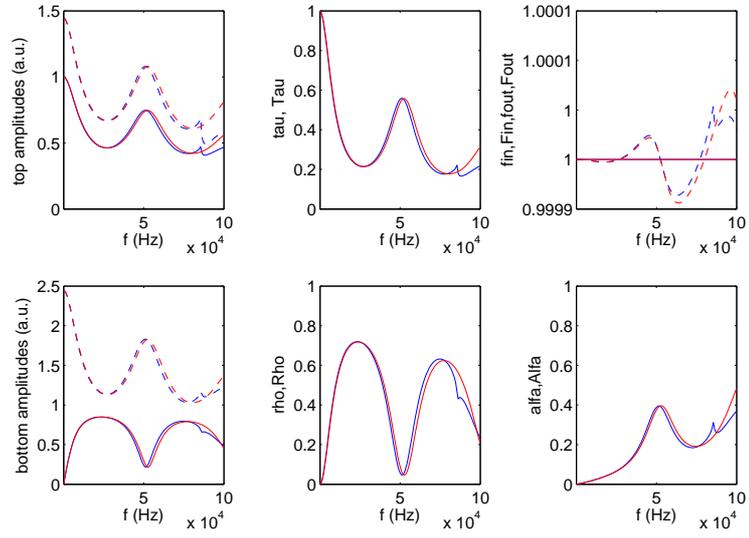}
 \caption{$h=0.00475 m$}
  \label{h-60}
  \end{center}
\end{figure}

\begin{figure}[ht]
\begin{center}
\includegraphics[width=12cm] {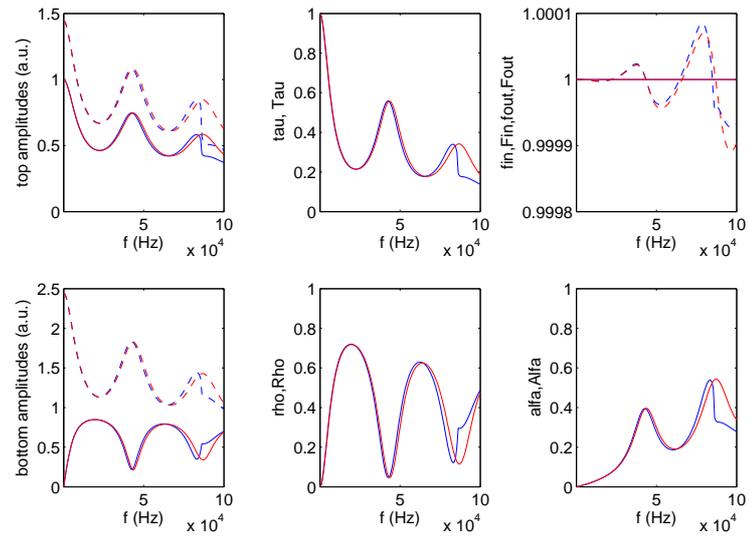}
 \caption{$h=0.00575 m$}
  \label{h-70}
  \end{center}
\end{figure}

\begin{figure}[ht]
\begin{center}
\includegraphics[width=12cm] {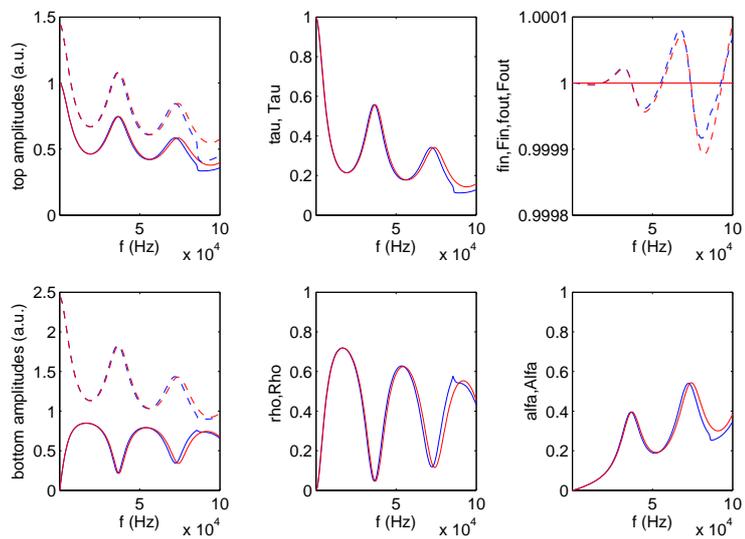}
 \caption{$h=0.00675 m$}
  \label{h-80}
  \end{center}
\end{figure}

In figs. \ref{h-10}-\ref{h-80}   we consider the case $\theta^{i}=0^{\circ}$, $w=0.003~m$, $C^{''[1]}=c^{''[1]}=-25~ms^{-1}$ for various $h$.
\clearpage
\newpage
These figures show that the agreement between the two responses is excellent, for all $h$, and for all frequencies up to the one at which the Wood anomaly occurs, this frequency being practically constant as a function of $h$. Of particular interest is the absorbed flux, which is practically the same for the grating and equivalent layer, whose maximum peak gradually shifts to lower frequencies and levels off in height beyond $h=0.00375~m$ while increasing in height, even with the formation of a second peak, starting at $h=0.00575~m$. Thus, the deep gratings (and equivalent layers) turn out to be quite efficient devices for wide-band absorption of acoustic waves.

A last observation concerns the near-perfect conservation of flux for both the grating and equivalent layer for all $h$.
\section{Conclusions}
The principal object of this investigation was to: (a)  obtain a simple model of acoustic response of block-like transmission gratings by an approximation procedure applied to a rigorous model of this response (b) evaluate the applicability and precision of approximate responses derived from the simple model by comparing them to the  numerical solutions obtained from the rigorous model.

It was shown that the approximate model amounts to considering the grating to behave, with respect to an acoustic solicitation, as a homogeneous layer of the same thickness as that of the blocks of the grating, the real and imaginary parts of the bulk wave velocity of this layer being equal to the real and imaginary parts of the bulk wave velocity in the interstitial medium of the   grating, and the mass density of the layer being equal to that of the interstitial medium divided by the filling factor ($\phi=w/d$) of the grating, all other constitutive parameters as well as the solicitation, being the same in the layer and grating configurations.

It was found, via a series of numerical tests, backed up by conservation of flux checks , that the  layer model  enables to predict many (but not all of) the principal features of seismic response for a wide range of grating thicknesses  provided the $\phi$ is  close to one (its maximum value) and the frequency of the solicitation is relatively low, these being the conditions invoked in the approximation procedure adopted to derive the (effective) layer response from that of the (rigorous) grating response.

It was found that a surprising feature of the layer model is  that it enables to predict the first (non-zero frequency) peak of EAT ('Extraordinary Acoustic Transmission' \cite{ll07},  which tranforms to significant absorption in the interstices when the latter are filled with a lossy medium), even for small $\phi$. However, the second peak, which is linked to the Wood anomaly, is not accounted-for in the layer model. In fact, the existence of the Wood anomalies (which occur for frequencies and incident angles at which an inhomogeneous scattered wave becomes homogeneous) is so tightly linked with the $d$-periodic nature of the scattering configuration (the inhomogeneous waves being absent in the layer modeel) that they cannot make their appearance in the response of a homogeneous layer unless the effective mass density and/or velocity of the surrogate layer are dispersive (this possibility was ruled out a priori herein, but was taken into accout in studies such as \cite{fb05,wi18}).

In spite of this shortcoming, the layer model deriving from our low-frequency homogenization scheme, appears to give meaningful predictions of the response of the transmission grating well beyond the static limit and can therefore be qualified as 'dynamical'. Moreover, these predictions can be improved either by the technique outlined in \cite{wi16c,wi18} or by taking into account higher-order iterates in the scheme presented herein. Finally, our  homogenization scheme may provide  a useful alternative to traditional multiscale and field averaging approaches to homogenization of periodic structures as regards their response to dynamic solicitations.
%

\end{document}